\documentclass[fleqn,usenatbib]{mnras}
\usepackage{newtxtext,newtxmath}
\usepackage[T1]{fontenc}
\usepackage{graphicx}	
\usepackage{amsmath}	
\usepackage{amssymb}	
\usepackage[version=4]{mhchem}
\usepackage{multirow}
\usepackage[english]{babel}
\usepackage{blindtext}

\newcommand{\mj}{M$_\mathrm{J}$}
\newcommand{\msol}{M$_\mathrm{\odot}$}
\newcommand{\rsol}{R$_\mathrm{\odot}$}
\newcommand{\racc}{$r_\mathrm{acc}$}
\newcommand{\rhill}{$r_\mathrm{H}$}

\title[Observing protoplanetary discs with the SKA - I]{Observing protoplanetary discs with the Square Kilometre Array - I. Characterising pebble substructure caused by forming planets}

\author[Ilee et al.]{John~D.~Ilee$^{1}$\thanks{E-mail: j.d.ilee@leeds.ac.uk},
Cassandra Hall$^{2,3,4}$,
Catherine Walsh$^{1}$,
Izaskun Jim\'enez-Serra$^{5}$,
\newauthor
Christophe Pinte$^{6,7}$,
Jason Terry$^{3,4}$,
Tyler Bourke$^{8}$ and
Melvin Hoare$^{1}$
\vspace{0.3em}
\\
$^1$School of Physics and Astronomy, University of Leeds, Leeds LS2 9JT, UK\\
$^{2}$Department of Physics and Astronomy, University of Leicester, University Road, Leicester, LE1 7RH, UK\\
$^{3}$Department of Physics and Astronomy, The University of Georgia, Athens, GA 30602, USA. \\
$^{4}$Center for Simulational Physics, The University of Georgia, Athens, GA 30602, USA.\\ 
$^{5}$Departamento de Astrof\'isica, Centro de Astrobiolog\'ia (CSIC-INTA), Carretera de Torrej\'on a Ajalvir km 4, Torrej\'on de Ardoz, Madrid 28850, Spain \\
$^{6}$School of Physics and Astronomy, Monash University, Clayton Vic 3800, Australia\\
$^{7}$Univ. Grenoble Alpes, CNRS, IPAG, F-38000 Grenoble, France\\
$^{8}$SKA Organisation, Jodrell Bank, Lower Withington, Macclesfield, Cheshire SK11 9FT, UK
}

\date{Accepted 2020 September 01. Received 2020 September 01; in original form 2020 August 19}

\pubyear{2020}

\begin{document}
\label{firstpage}
\pagerange{\pageref{firstpage}--\pageref{lastpage}}
\maketitle

\begin{abstract}
High angular resolution observations of discs at mm wavelengths (on scales of a few au) are now commonplace, but there is a current lack of a comparable angular resolution for observations at cm wavelengths. This presents a significant barrier to improving our understanding of planet formation, in particular how dust grains grow from mm to cm sizes. In this paper, we examine the ability of the Square Kilometre Array (SKA) to observe dust substructure in a young, planet-forming disc at cm wavelengths.  We use dusty hydrodynamics and continuum radiative transfer to predict the distribution and emission of 1\,cm dust grains (or pebbles) within the disc, and simulate continuum observations with the current SKA1-MID design baseline at frequencies of 12.5\,GHz (Band 5b, $\sim$2.4\,cm) on 5--10\,au scales.   The SKA will provide high-fidelity observations of the cm dust emission substructure in discs for integration times totalling 100's of hours. Radial structure can be obtained at a sufficient resolution and S/N from shorter (10's of hours) integration times by azimuthal averaging in the image plane. By modelling the intensity distribution directly in the visibility plane, it is possible to recover a similar level of (axisymmetric) structural detail from observations with integration times 1--2 orders of magnitude lower than required for high-fidelity imaging. Our results demonstrate that SKA1-MID will provide crucial constraints on the distribution and morphology of the raw material for building planets, the pebbles in protoplanetary discs.  
\end{abstract}

\begin{keywords}
stars: pre-main-sequence -- protoplanetary discs -- planet–disc interactions -- planets and satellites: formation -- radio continuum: planetary systems
\end{keywords}



\section{Introduction}
\label{sec:intro}

Thanks to the Atacama Large Millimetre Array (ALMA), it has become clear that several protoplanetary discs possess a common physical structure in the form of concentric dust rings and gaps \citep[e.g.][]{almapartnership2015, fedele_2017, dipierroetal2018, clarke_2018, Long2019}.  The origin of these dust structures has been a matter of debate, with possibilities ranging from dust opacity changes at molecular ice lines \citep{Zhang2015,Okuzumi2016} to on-going (magneto-)hydrodynamic instabilities within the disc \citep[e.g.,][]{Flock2015,Suriano2018,Riols2019}.  While some of these possibilities have been suggested to be unlikely for deep and wide gaps \citep[for example ice lines, see e.g.][]{huang_2018, vanderMarel2019}, a popular explanation for such structures is the presence of still-forming planets in the disc \citep{Paardekooper2004,Dong2015}.

\smallskip

This interpretation has been strengthened in recent years by the direct detection of potential signatures of young planets, including localised continuum excesses \citep[][]{keppler_2018, tsukagoshi_2019,nayakshinetal2020} and kinematic deviations in the molecular gas of the disc caused by a perturbing planet \citep[e.g.][]{perezetal2015,teague_2018, pinte_2019, pinte_2020}.  Perhaps the strongest evidence comes not from observations of discs, but from direct imaging of mature planetary systems.  Many such observational campaigns have revealed the presence of bona-fide planets at comparable distances to their stars as these disc substructures \citep[e.g.][]{marois_2008, lagrange_2010, vigan_2017}. This demonstrates that (at least part of) the observed exoplanet population may have been formed in protoplanetary discs at the location of observed rings and gaps.

\smallskip

Beyond simply inferring the presence of planets, a number of works have also shown that disc substructures can be used to derive further information on the disc and forming planets.  The width and depth of gaps in the dust (and the gas) surface density can be used to place constraints on the mass of the planets that are opening them \citep[e.g.][]{crida_2006, kanagawa_2015, zhang_2018}.  Single planets have been shown to open multiple gaps if the viscosity is low \citep{dong_2018, fedele_2018} or if they are migrating sufficiently quickly \citep{meru_2019, nazari_2019}. It is now also understood that observations of substructure caused by forming planets at multiple wavelengths may offer a way to constrain disc masses \citep{veronesietal2019}.  Finally, the width dust continuum emission rings may be correlated with the strength of turbulence in the disc \citep{dullemond_2018}.

\smallskip

Despite the large amount of observational evidence of planet-induced substructure in discs, there is still significant debate over \emph{how} planets form within discs.  A major issue arises from the fact that typical disc lifetimes \citep[$\lesssim$10\,Myr,][]{ribas_2014} appear to be shorter than the timescales required to assemble giant planets at large radii ($\gtrsim$10\,au).  The classical oligarchic growth model \citep{safronov_1967} involves the collision and sticking of dust particles to form progressively larger bodies in the disc.  However, the growth of millimetre-sized dust to kilometre-sized planetesimals is difficult to explain since a variety of barriers exist. Collisions of millimetre-sized particles tend to result in either fragmentation or bouncing rather than coalescence \citep{blum_2008}.  In addition, solid material in the disc can experience aerodynamic drag from gas which removes angular momentum, causing rapid inward drift towards the central star \citep{weidenschilling_1977}.  While several processes have been proposed that enable rapid growth (such as the streaming instability, see \citealt{youdin_2005} and \citealt{johansen_2007}), it is clear that the process of growth from millimetre-sized dust grains to centimetre-sized \emph{pebbles} needs to be better understood in order to fully decipher the planet formation process.  At millimetre wavelengths, dust opacity is dominated by millimetre-sized dust grains \citep[see, e.g.,][]{drainelee1984, semenov_2003, draine_2006}.  Therefore ALMA, which covers the wavelength range between 3\,mm up to 0.3\,mm, can observe emission from approximately millimetre-sized dust grains.  However, ALMA is not sensitive to emission from dust grains of centimetre-sizes and larger.  Therefore, the best way to probe the growth and evolution of dust grains into this regime is to carry out observations of the dust thermal emission at longer (cm) wavelengths. 

\smallskip

First steps toward this goal have involved combining observations at mm-wavelengths from instruments such as the Submillimeter Array (SMA) and ALMA, with cm-observations using the Karl G.\ Janksy Very Large Array (JVLA).  For example, \citet{perez_2012} found evidence that cm-wavelength emission from the disc in the AS~209 system is significantly more compact than in the millimetre, suggesting a bulk radial variation in dust opacity that can be explained by particle growth and subsequent drift to the inner disc.  Further studies (including results from large observing programs such as the Disks@EVLA project) have revealed similar radial variations in opacity across many discs that can be explained by the growth and drift of dust grains (e.g. CY~Tau, DoAr~25; \citealt{perez_2015}; HD~163296; \citealt{guidi_2016}; AS 209, FT Tau, DR Tau; \citealt{tazzari_2016}).  More recently, \citet{carrasco-gonzalez_2019} performed an extensive multi-wavelength study of the HL~Tau protoplanetary disc using observations at 0.9, 1.3, 2.1, 7, 9 and 13\,mm, enabling a radially-resolved determination of changes in dust grain sizes up to a few millimetres. Such radially-resolved characterisation will be essential to observationally verify theories of dust growth across the cm-size barrier, but this is not possible to achieve with current instrumentation due to the limited spatial resolution and sensitivity available.

\smallskip

The Square Kilometre Array (SKA) will soon revolutionise interferometric observations at cm-wavelengths \citep[see][]{braun_2015}. The mid-frequency array, SKA1-MID, will consist of 133 dishes spread across the Karoo desert in the Northern Cape Province of South Africa.  With maximum baselines of the order of 150\,km, SKA1-MID will provide extremely high resolution observations at frequencies between 350\,MHz -- 15.3\,GHz.  Within the proposed Band 5b frequency range (8.3--15.3\,GHz, 3.6--2.0\,cm), angular resolutions down to $\sim$0$\farcs$04 will be possible \citep{braun_2019}.  These correspond to scales of $\sim$5\,au for the nearest star forming regions at distances of 140\,pc, which is a factor 4--5 better than the highest angular resolutions available with the current JVLA configurations at the same frequencies.  In addition, the extremely large collecting area of SKA1 will result in sensitivities that are a factor of 3--4 higher than those currently available with the JVLA dishes.  Several works have examined the potential that the SKA has in providing significant contributions in our understanding of star and planet formation \citep{hoare_2015, testi_2015}, and pathfinder studies using the JVLA have started to determine optimal targets for first observations \citep{coutens_2019}.  However, there are currently no detailed investigations on the potential for the SKA to characterise the cm-dust component of planet forming discs.  

\smallskip

In this paper, we use a combination of dusty hydrodynamic simulations coupled with a continuum radiative transfer code in order to predict the morphology of cm-sized dust grains in a young protoplanetary disc hosting forming planets.  We predict the observed sky brightness distribution of the model at frequencies appropriate for Band 5b of the SKA, and simulate observations of the disc at 12.5\,GHz ($\sim$2.4\,cm) using the proposed array configuration and sensitivity limits.  We examine the detectability of the planet-induced substructure with a variety of integration times and observing configurations, and outline several data analysis techniques that enable important physical parameters to be extracted from the observations.  The paper is structured as follows -- Section \ref{sec:methods} outlines our methodology, Sections \ref{sec:results} \& \ref{sec:analysis} present our results and summarises our analysis, we discuss these in context in Section \ref{sec:discussion}, and finally Section \ref{sec:conclusions} presents our conclusions.

\section{Methods}
\label{sec:methods}

\begin{figure*}
\centering
\includegraphics[width=\textwidth]{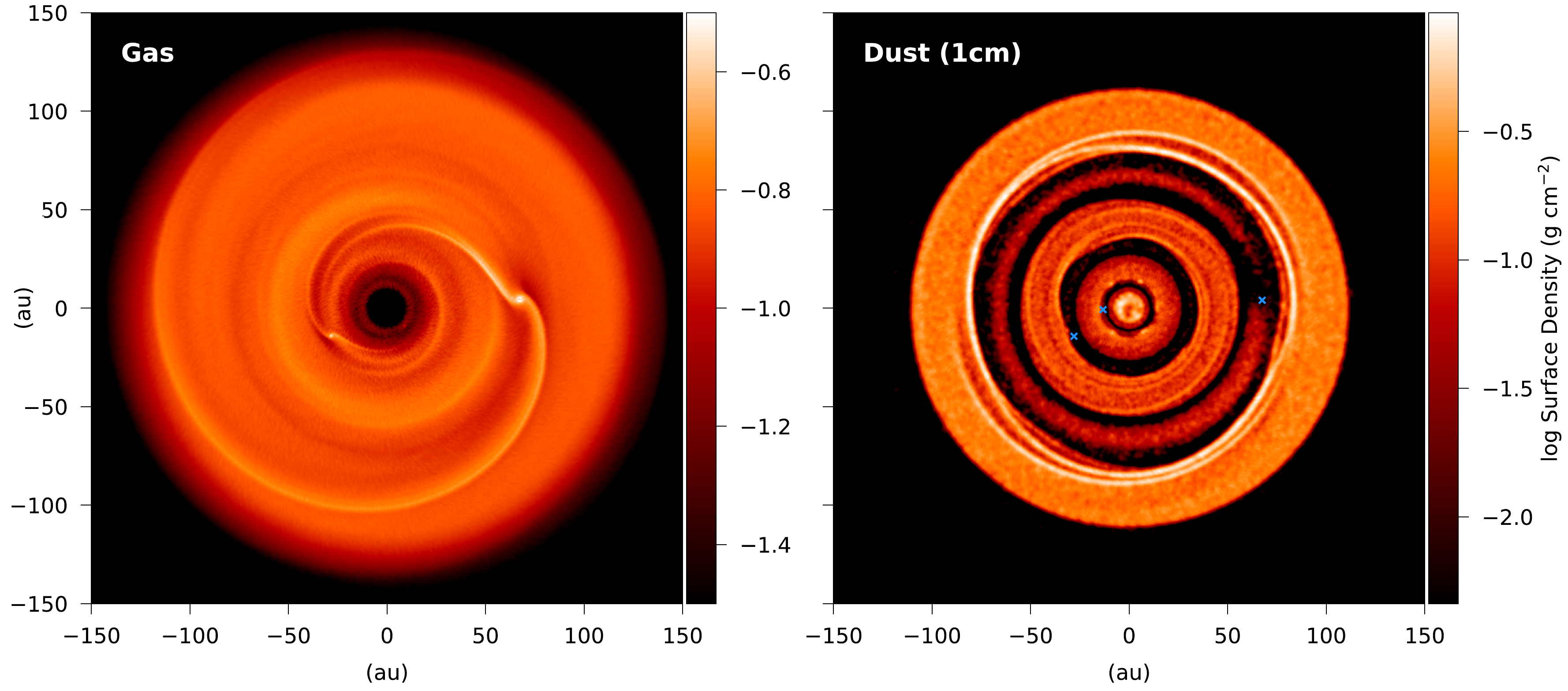}
\caption{Surface density of gas (left) and dust grains of size $a=1$\,cm (right) in the hydrodynamic model of the protoplanetary disc after 12,000\,yrs of evolution.  The location of the planets are marked with blue crosses in the right panel.}
\label{fig:hydro}
\end{figure*}

\subsection{Dusty hydrodynamics simulations} 

In this study, we use a generic disc model that is not intended to be representative of any particular source.  However, in order to ensure our results are representative of real systems (and to keep within a manageable parameter space), we emulate the disc structure and (possible) planetary architecture of HL~Tau star-disc system following \cite{almapartnership2015} and \citet{dipierro_2015_hltau}, which we outline below.

\smallskip

We use the three-dimensional smoothed particle hydrodynamics (SPH) code \texttt{PHANTOM} to model both a gas and dust component \citep{phantom}.  For the dust component, we consider a single population of grains of radius $a=1$\,cm with a bulk intrinsic grain density of 1\,g\,cm$^{-3}$ \citep[in broad agreement with the recent results of ][]{birnstiel_2018}.  At this size, the grains are large enough to be in the regime of weak drag with respect to the gas (i.e.\ large Stokes numbers), and so we utilise the \emph{two-fluid} approach treating the dust and gas as interacting fluids as described in \citet{laibeprice2012a,laibeprice2012b}. Each simulation is evolved with $10^6$ SPH particles, and $2.5\times 10^5$ dust particles, which avoids dust becoming artificially trapped below the gas resolution \cite[see][]{laibeprice2012b}.

\subsubsection{Disc mass and gas-to-dust ratio}

It is now largely understood that the dust-to-gas mass ratio $\epsilon$ in protoplanetary discs is unlikely to match the canonical value in the interstellar medium of $0.01$ \citep{mathisetal1977}. For example, recent observations of CO isotopologues and dust continuum emission have found disc-averaged $\epsilon$ values to be much higher ($\sim$0.2, see \citealt{ansdelletal2016}), and simulations have demonstrated that discs can be formed with significant dust enrichment \cite{lebreuilly_2020}.  There is also an observed discrepancy between the radial extents of the gas and dust components of protoplanetary discs, with dust discs being significantly more compact \citep{Andrews2012,perez_2012,perez_2015}. In some cases, such as IM~Lup, the gas disc may be 10 times larger in radial extent than the mm dust disc \citep[e.g.][]{cleevesetal2016}.  This discrepancy can explained by a combination of grain growth and radial drift. Essentially, the radial velocity, $v_\mathrm{r}$, of dust particles has two components: drag, $v_\mathrm{drag}$, and drift, $v_\mathrm{drift}$, and is given by 
\begin{equation}
\label{eq:drift}
v_{\mathrm{r}}=v_{\mathrm{drag}}+v_{\mathrm{drift}}=\frac{v_{\mathrm{g}, \mathrm{r}}}{1+{St}^{2}}+\frac{1}{{St}+{St}^{-1}} \frac{1}{\rho_{\mathrm{g}} \Omega} \frac{\partial P}{\partial R},
\end{equation}
where $v_\mathrm{g,r}$ is the radial velocity of the gas, $\rho_\mathrm{g}$ is the gas mass surface density, $\Omega$ is the Keplerian angular velocity and $P$ is the pressure \citep{weidenschilling_1977}. 
The Stokes number, $St$, determines the degree of coupling to the gas.  For a vertically isothermal disc (as used here) this is expressed simply as
\begin{equation}
\label{eq:stokes}
    St = \frac{\pi a\rho_\mathrm{s}}{2\Sigma_\mathrm{g}}, 
\end{equation}
where $a$ is grain size, $\rho_\mathrm{s}$ is intrinsic bulk grain density and $\Sigma_\mathrm{g}$ is the gas surface density. Examining equations \ref{eq:drift} and \ref{eq:stokes} , we can see that the drag term decreases as grain size increases. However, the drift term peaks at $St=1$, which means that the grain size for which radial drift is fastest depends on the surface density, and therefore location, within the disc. 

\smallskip

This provides us with a simple physical argument that explains the increased dust-to-gas ratio of the disc observed in the dust continuum -- radial drift. To work out exactly how much enhancement takes place, we assume that total dust mass is conserved, and shortly after formation, before any radial drift has occurred, that the dust mass $M_\mathrm{dust}$ is expressed as
\begin{equation}
    M_\mathrm{dust} = \epsilon_{\mathrm{ISM}}\Sigma_\mathrm{g} \pi R_\mathrm{gas\,disc}^2.
\end{equation}
Given that $M_\mathrm{dust}$ is conserved, we can equate this expression with one that describes the dust after radial migration has taken place, and the dust disc is now more compact than the gas disc,
\begin{equation}
    \epsilon_{\mathrm{ISM}}\Sigma_\mathrm{g} \pi R_\mathrm{gas\,disc}^2 = \epsilon_{\mathrm{enhanced}}\Sigma_\mathrm{g} \pi R_\mathrm{dust\,disc}^2,
\end{equation}
which reduces to 
\begin{equation}
    \epsilon_{\mathrm{enhanced}} = \epsilon_{\mathrm{ISM}} \Bigg[\frac{R_\mathrm{gas\,disc}}{R_\mathrm{dust\,disc}}\Bigg]^2.
\end{equation}

As discussed earlier, we know that $R_\mathrm{gas\,disc}/R_\mathrm{dust\,disc}\sim 10$, and arrive at $ \epsilon_{\mathrm{enhanced}} = 1.0$ within 120\,au.  We note that this is likely to be an upper limit.  We therefore adopt $M_{\rm gas} = M_{\rm dust} = 0.001$\,\msol for the hydrodynamic modelling, which is consistent with previous studies of the formation of dust substructure in protoplanetary discs \citep[e.g.][]{dipierroetal2018,pinte_2019}.

\subsubsection{Disc structure \& initial conditions}

The central star and planets are represented by sink particles \citep{bate1995}, with planets initially located at radii of 13.2, 32.3 and 68.8\,au with masses of 0.20\,\mj, 0.27\,\mj, and 0.55\,\mj, respectively.  We set the accretion radius, $r_\mathrm{acc}$ of the central star as \racc = 1\,au, and the accretion radius of the respective planets as fractions of their respective Hill radii, such that \racc = 0.5\,\rhill, 0.25\,\rhill and 0.2\,\rhill. For a circular orbit, the Hill radius is given by:
\begin{equation}
    r_\mathrm{H} \approx a_{\rm maj} \bigg(\frac{m_\mathrm{p}}{3M_\mathrm{*}}\bigg),
\end{equation}
where $a_{\rm maj}$ is semi-major axis, $m_\mathrm{p}$ is planet mass and $M_*$ is central stellar mass.  For this system, these correspond to accretion radii of 0.24\,au, 0.32\,au and 0.71\,au respectively. Within these radii, material is able to freely accrete on to the sink particle as long as it is gravitationally bound to the sink and has a negative velocity divergence.  The location of the central star is not fixed and the planets are free to migrate through the disc, although we note that over the timescales that we consider this migration is negligible (less than $\lesssim$0.5\,au in all cases).  

\smallskip

We set the disc inner radius to $R_\mathrm{in} = 1$ au and the outer disc radius $R_\mathrm{out} = 120$ au. We assume that the disc surface density profile is a power law given by $\Sigma(R) = \Sigma_0 (R/R_\mathrm{in})^{-p} $, where $\Sigma_0$ is the disc surface density at $R_\mathrm{in}$. We assume that the disc has a locally isothermal equation of state such that the pressure $ P = c_\mathrm{s}^2(R)\rho$, where $c_\mathrm{s}$ is the sound speed and $\rho$ is the mass density. The sound speed profile obeys $c_\mathrm{s} = c_\mathrm{s,0} (R/R_\mathrm{in})^{-q}$, where $c_\mathrm{s,0}$ is the sound speed at $R=R_\mathrm{in}$. The gas temperature is then given by $T(R) \propto (R/R_\mathrm{in})^{-2q}$, with aspect ratio $H/R = (H/R)_\mathrm{in} (R/R_\mathrm{in})^{\frac{1}{2}-q}$.  \citet{dipierro_2015_hltau} constrained the aspect ratio of the disc from ALMA observations \citep{almapartnership2015}, and determined that $T\propto R^{-0.7}$, which gives $q=0.35$. We choose a shallow surface density profile such that $p=0.1$ for computational efficiency, but note that this may result in excess mass in the outer parts of the disc. We set $(H/R)_\mathrm{in} = 0.04$, again matching the constraints in \citet{dipierro_2015_hltau}. We set the SPH artificial viscosity $\alpha_{\mathrm{SPH}} = 0.13$, which gives an effective  $\alpha_{\mathrm{SS}} \sim 0.005$ \citep{ss1973}. The simulation is run for approximately 10 orbits at the disc outer edge (12000\,yrs), which corresponds to approximately 20 orbital periods for the outermost planet.  Figure \ref{fig:hydro} shows the resulting surface density of the gas and dust components of the disc at the end of the simulation.

\begin{figure}
\includegraphics[width=\columnwidth]{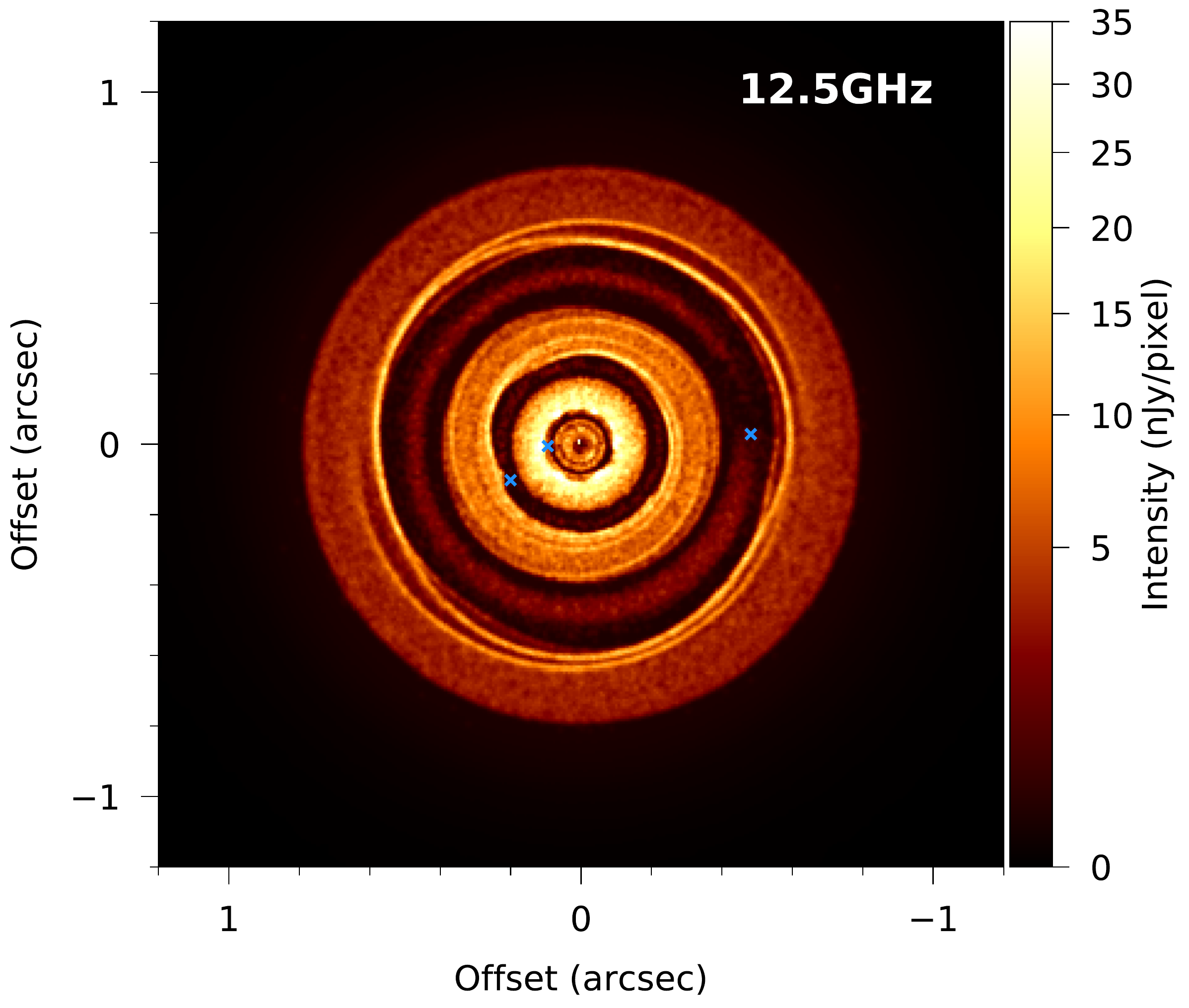}
\caption{Resulting intensity from the radiative transfer calculations at a frequency of 12.5\,GHz.  The locations of the planets are shown with blue crosses, and the colour scale is shown with square-root stretch in units of nJy / pixel, where a pixel is 7\,mas in size.}
\label{fig:RT}
\end{figure}

\begin{figure*}
\includegraphics[width=\textwidth]{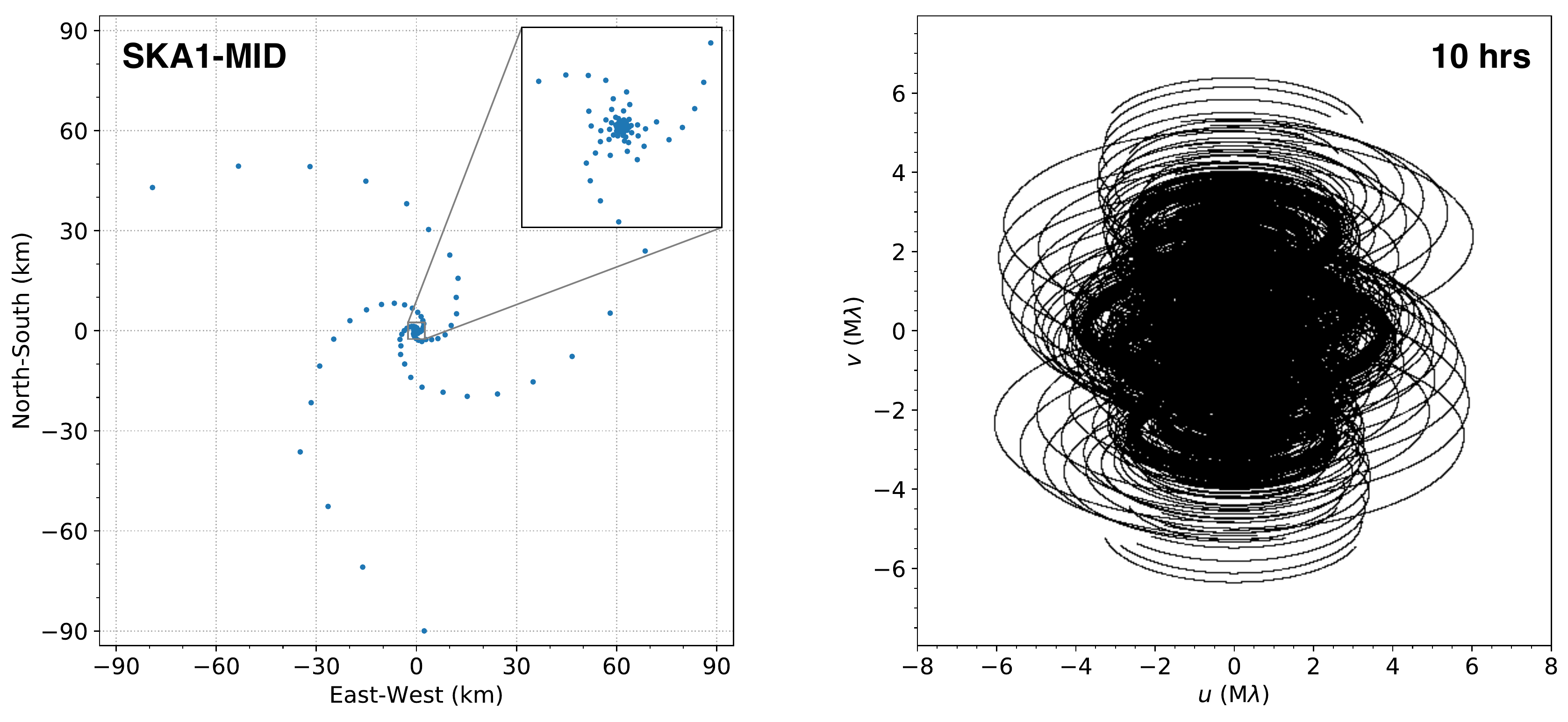}
\caption{SKA1-MID antenna positions assumed for the simulated observations (left).  The array consists of 133 dishes with a maximum baseline of 154\,km.  The inset shows the inner $5\times5$\,km and the core of the array. Corresponding $uv$-coverage obtained with the assumed antenna configuration for a continuous integration time of 10\,hrs (right).}
\label{fig:ska_array}
\end{figure*}

\subsection{Radiative transfer calculations}
\label{sec:mcfost}

Armed with the \emph{morphology} of the cm-sized dust within the protoplanetary system described above, we then utilise the Monte Carlo radiative transfer code MCFOST \citep{mcfost} to compute the disc thermal structure and observed appearance.  We assume that the gas and dust temperatures are in thermal equilibrium (e.g. $T_\mathrm{gas} = T_\mathrm{dust}$, which is justified due to the high densities expected in disc midplanes), and calculate $T_\mathrm{dust}$ using $10^8$ photon packets.  The central star is assumed to have a mass of 2\,\msol\ with a radius of 1.48\,\rsol\ and effective temperature of $T_\mathrm{eff}=10,600$\,K.  The star-disc system was assumed to lie at a distance of 140\,pc, and is viewed at an inclination of zero degrees (i.e.\ a face-on orientation).  The density structure of the SPH simulation underwent Voronoi tesselation such that each SPH particle corresponds to an MCFOST cell. The dust composition was set to a mixture of amorphous carbon and silicates \citep{drainelee1984}, with optical properties calculated using Mie theory. 

\smallskip

Within each Voronoi cell, the distribution of dust grain sizes is split into 100 logarithmic bins from $a_\mathrm{min}=0.1$\,$\mu$m to $a_\mathrm{max}=2$\,cm.  Grains smaller than 1\,$\mu$m are assumed to perfectly trace the gas, and those larger are interpolated between the gas and 1\,cm dust.  We assume a power-law relationship between grain size $a$ and number density of dust grains $n(a)$ such that d$n(a)\propto a^{-3}$\,d$a$, which is appropriate for a modest amount of grain growth.  This size distribution is then normalised such that the assumed total dust mass is increased to 0.01\msol\footnote{Since the morphology of the dust depends largely on $St$, and $St$ has no dependence on dust density (other than through intrinsic bulk density), then increasing or decreasing the total dust mass present does not affect the morphology of the dust.}, in agreement with recent observations suggesting a higher disc mass for HL~Tau \citep[see, e.g.,][]{tapia_2019, carrasco-gonzalez_2019, booth_2020}.  The diaelectric indices for each grain size were tabulated values from \citet{drainelee1984} from 0.1\,$\mu$m to 1000\,$\mu$m. Beyond this, opacities were extrapolated linearly in log-log space to the observing frequency of 12.5\,GHz, leading to $\kappa = 0.63$\,cm$^{2}$\,g$^{-1}$.  The resulting on-sky surface brightness distribution is shown in Figure \ref{fig:RT}, and results in a total disc integrated flux density of 0.15\,mJy, which is comparable to the free-free corrected measurements of dust continuum emission from young discs at similar frequencies (e.g.\ HL~Tau, \citealt{carrasco-gonzalez_2019}; the $\rho$~Oph~A members, \citealt{coutens_2019}).

\subsection{Simulating observations with SKA1-MID}

We assume that SKA1-MID will consist of 133 dishes of 15\,m diameter arranged in a spiral configuration as show in Figure \ref{fig:ska_array} (left).  The array is assumed to be centred at a latitude of 30\degr42\arcmin46\farcs53\,South and longitude of 21\degr26\arcmin 37\farcs6908\,East.  This configuration results in maximum and minimum baselines of 154\,km and 15\,m, respectively.  In this work, our observations are assumed to be performed at 12.5\,GHz, which is at the centre of the proposed Band 5 frequency range (8.3--15.3\,GHz). Our target is assumed to lie within the Ophiuchus A star forming region at $16^{h}26^{m}30^{s}$, $-24^{d}24^{m}0^{s}$ (which is close to the middle of the accessible declination range for the SKA).  Our sensitivity estimates assume total continuum bandwidth of 5\,GHz (with a fractional bandwidth of $\Delta \nu / \nu = 0.3$), and are detailed in Table \ref{tab:sens}.  We consider two imaging scenarios -- the first has an angular resolution of $\theta = 0.034''$ and a sensitivity of $\sigma_{\rm c} = 2.4$\,$\mu$Jy/beam, while the second has an angular resolution of $\theta = 0.067''$ and a sensitivity of $\sigma_{\rm c} = 1.2$\,$\mu$Jy/beam (where sensitivities are quoted for an integration time of one hour, and angular resolutions are set by tapering the visibilities). Using the \texttt{simulator} toolkit\footnote{\url{https://casa.nrao.edu/casadocs/casa-5.1.1/global-tool-list/tool_simulator}} in CASA 5.1.2 \citep{mcmullin_2007}, the on-sky intensity of the radiative transfer images are Fourier transformed into the visibility (or $uv$) plane, the coverage of which can be seen in Figure \ref{fig:ska_array} (right) for a continuous observation time of 10\,hrs.  The real and imaginary parts of the visibility are then corrupted with thermal noise ($\sigma_{\rm v}$) at a level appropriate for the assumed angular resolution and sensitivity of the observations (see Table \ref{tab:sens}).  Synthetic images are then reconstructed by utilising a uniform weighting scheme with a Gaussian taper to the visibility weights in order to recover the desired angular resolution.  For a full discussion of the anticipated sensitivity and performance of the SKA1 arrays, see \cite{braun_2019}.

\begin{table}
 \centering
 \caption{Angular resolutions, sensitivities and effective visibility noise assumed during the calculation of the synthetic observations, following \citet{braun_2019}.  The sensitivity and effective visibility noise are applicable to an observation with an integration time of 1 hour.}
 \begin{tabular}{cccc}
  \hline
  Frequency     & Angular Resolution            & Sensitivity                  & Visibility Noise     \\
      $\nu$          &   $\theta$        &  $\sigma_{\rm c}$                      &  $\sigma_{\rm v}$                \\
   (GHz)   &  (mas)    &    ($\mu$Jy/beam)    &  (mJy)            \\
  \hline
  12.5  &   34   &  2.4     & 0.56   \\   
  12.5  &   67   &  1.2     & 0.28   \\ 
  \hline
 \end{tabular}
 \label{tab:sens}
\end{table}

\section{Results}
\label{sec:results}

\subsection{Morphology of the dust and gas in the SPH simulations}

Examining Figure \ref{fig:hydro}, there is a very different morphology for the surface density of gas and 1\,cm dust grains in the disc model.  The three planets at 13, 32 and 69\,au are not massive enough (at 0.20\,\mj, 0.27\,\mj, and 0.55\,\mj, respectively) to open significant gaps in the gas disc, but more easily open gaps in the dust disc.  These morphological differences can be understood by the differing level of interaction between gas and dust based on the grain size.  Dust grains, if unimpeded by gas, would move in exact Keplerian rotation. However, gas in the disc is partially supported against gravity by an outwards pressure gradient, causing it to orbit at sub-Keplerian speed. This means that the dust experiences a headwind, and ultimately exchanges angular momentum with the gas and undergoes radial drift.  The amount of drift experienced by a grain is related to the size by the functional form described in Appendix \ref{sec:drift}.  The presence of planets in the disc alters the pressure gradient, leading to increased clearing of grains in the vicinity of the planet's orbit and a `pile-up' of grains at the outer edge of these gaps \citep[see also][]{pinilla_2015}, mostly easily seen toward at the outer edge of the outermost gap (80\,au). In addition to the general radial variations, there are a number of non-axisymmetric features that are apparent in the simulation.  In particular, the gaps in the dust distribution are not entirely empty, and are filled with `extended horseshoes' of grains, again most easily seen within the outermost gap (70\,au).  

\subsection{Morphology of the simulated SKA observations}

\begin{figure*}
\centering
\includegraphics[width=0.98\textwidth]{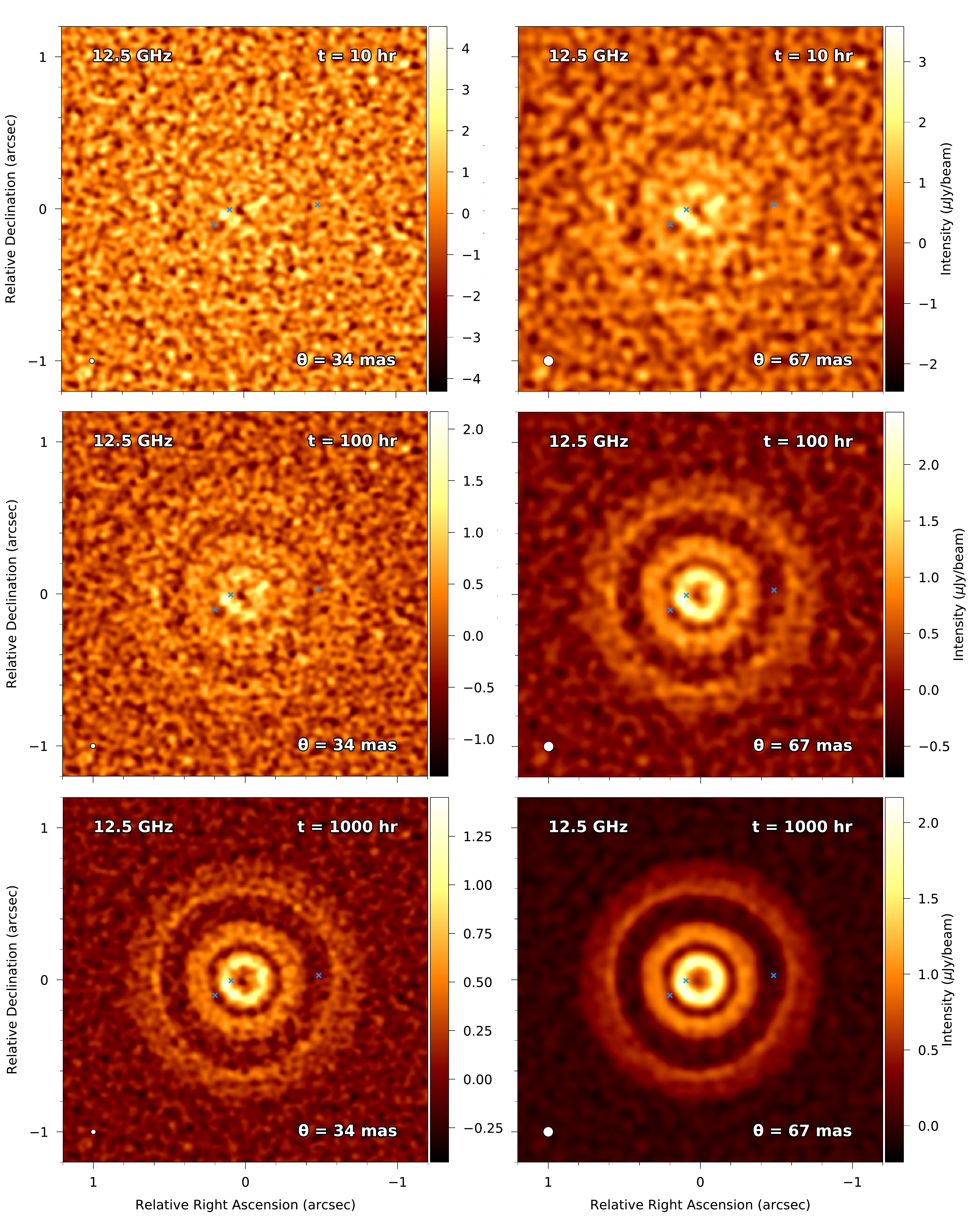}
\caption{Simulated continuum observations at 12.5\,GHz for angular resolutions of 34\,mas (left) and 67\,mas (right) and integration times of 10\,hrs (top), 100\,hrs (middle) and 1000\,hrs (bottom).  The positions of the planets in the model are shown with blue crosses. }
\label{fig:all_obs}
\end{figure*}

\begin{table}
 \centering
 \caption{Observational parameters as measured from the synthetic images in Figure \ref{fig:all_obs}.}
 \begin{tabular}{ccccc}
 \hline  
Integration  & \multicolumn{2}{c}{$\sigma_{\rm rms}$ ($\mu$Jy/beam)} & \multicolumn{2}{c}{Peak SNR}\\       
(hrs)             & 34\,mas & 67\,mas & 34\,mas & 67\,mas \\
 \hline
1                 &  2.64 & 1.39 & 4.16  & 4.52   \\ 
10                &  0.83 & 0.44 & 4.90  & 7.58   \\
100               &  0.26 & 0.14 & 7.73  & 17.17  \\
1000              &  0.08 & 0.05 & 16.8  & 44.75  \\
  \hline
 \end{tabular}
 \label{tab:obs}
\end{table}

Figure \ref{fig:all_obs} presents the synthetic SKA observations of the model sky brightness distribution.  We examine observations made with the smallest possible angular resolution (34\,mas) and a 1\,hr sensitivity of 2.4\,$\mu$Jy/beam, along with a slightly lower angular resolution (67\,mas) at a 1\,hr sensitivity of 1.2\,$\mu$Jy/beam.  The choice of these settings is motivated by the desire to test whether the highest angular resolution is required to accurately characterise disc substructure, or if higher sensitivity is more important in its recovery.  For each of these, we investigate four integration times of 1, 10, 100 and 1000\,hrs.  We have omitted the 1\,hr observations from Fig \ref{fig:all_obs} for clarity since none of these produced any observable features.  

\smallskip

For observations with $\theta=34$\,mas, the structure of the disc is only apparent in the images with 100 and 1000\,hr integration times.  For the former, the detection of ringed structure ranges in significance from 3--6$\sigma$ from the outer to inner disc regions (where $\sigma$ is the corresponding rms level from Table \ref{tab:obs}).  For the latter, the outer disc is detected at between 6--7$\sigma$ and in the inner disc at 16$\sigma$.   Of particular note is the fact that the 1000\,hr integration image is able to directly resolve the `pile-up' of cm-sized grains at the inner edge of the outermost ring at $\sim$80\,au ($0\farcs6$).   

\smallskip

For observations with $\theta=67$\,mas, the disc structure is detectable in images with 10, 100 and 1000\,hr integration times at significances ranging from 3-8$\sigma$ for the 10\,hr image, up to 15--40$\sigma$ for the 1000\,hr image.   We note that while the particular hydrodynamic simulation we use here shows no significant azimuthal variations in this feature, images of this fidelity would be able to detect them if present (which would be difficult to discriminate from noise features in the $\theta=34$\,mas image).

\section{Analysis}
\label{sec:analysis}

In this section, we explore several techniques that can be applied to both the observed images and raw visibilities in order to detect (and characterise) any disc substructure.

\begin{figure}
\includegraphics[width=0.49\textwidth]{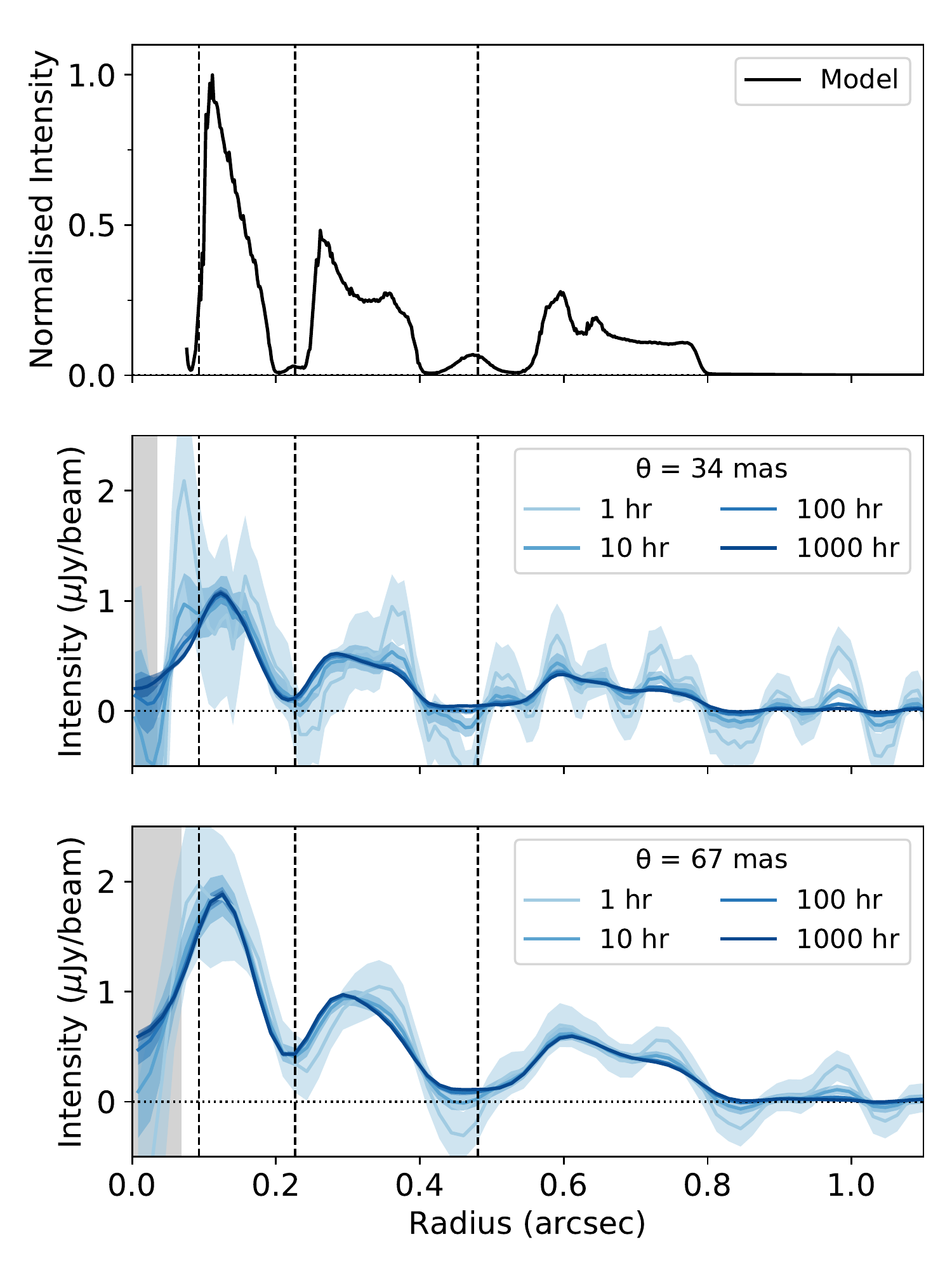}
\caption{Radial profiles calculated from the simulated observations.  The upper panels show the profile of the radiative transfer model (with the innermost region masked due to excess noise), the middle panel shows profiles for observations with a beam size of 34\,mas, and the lower panel shows those for a beam size of 67\,mas.  The shaded area indicates the standard deviation of each annulus, and line colours denote the assumed integration time for each observation (1, 10, 100 \& 1000\,hrs).  The radial locations of the planets in the hydrodynamic simulation are shown with dashed vertical lines.}
\label{fig:radial_image_profiles}
\end{figure}

\subsection{Intensity profiles in the image plane}

We performed azimuthal averaging of the observed images.  For each image, we constructed a grid from the centre of the image out to a radius of $1\farcs2$ ($\sim$170\,au) with radial resolution corresponding to one-third of the beam major axis.  For each radial bin, the pixel values are summed to produce the radial profile, and the uncertainties are calculated as the standard deviation divided by the square root of the number of beams along each annulus.  Radial profiles for each set of images are shown in Figure \ref{fig:radial_image_profiles} (along with a radial profile of the on-sky radiative transfer image for comparison).

\smallskip

The main result to note from the radial profiles is that the presence of gaps in the disc emission can be inferred even for short integration times of between 1--10\,hrs, particularly for images in which no discernible emission can be seen in Figure \ref{fig:all_obs}.  However, for the lower sensitivity observations with $\theta=34$mas and integration times of 1\,hr, it may be difficult to assign this structure to the location of gaps without knowing their position \emph{a priori}, since the peak-to-trough variations are of the order of the uncertainties on the radial profile itself.  

\smallskip

Beyond simply detecting the presence of a gap, the radial profiles also allow characterisation of the gap properties such as location and shape (e.g.\ width and/or depth).  While the location of a gap has the potential to reveal the location of any planet causing it, it has also been demonstrated that the gap shape can give information on the planet mass \citep[see e.g.][]{kanagawa_2015, rosotti_2016}.  Examining the profiles in Figure \ref{fig:radial_image_profiles}, it is clear that innermost gap (caused by the planet at 13\,au, $\sim$0$\farcs$1) cannot be accurately characterised by observations at these spatial resolutions.  In addition, the relatively flat profile of the outermost gap at 69\,au ($\sim$0$\farcs$5) means a retrieval of the gap centre is non-trivial.  We therefore concentrate on comparisons involving the recovery of the location of the middle gap caused by the planet located at a radius of 32\,au ($\sim$0$\farcs$2).  The radial profiles at the highest resolution of $\theta=34$\,mas only recover the location of the planet (shown with a vertical dotted line) to within $\sim$1\,au for integration times of 100--1000\,hrs, while observations at $\theta=67$\,mas recover the planet location to a similar accuracy after only 10\,hrs.

\smallskip

We note that the correct recovery of an accurate radial profile is predicated knowing the central location of the emission to a sufficient degree of accuracy, which would be difficult to determine from the low signal-to-noise images.  It would be possible to recover the central location via analysis of the imaginary component of the observed visibilities.   Minimisation of this component would act to recover the most azimuthally symmetric intensity distribution, which for a disc with little non-axisymmetric structure, would be close to the disc centre.  Alternatively, using complementary observations of the same source with higher signal-to-noise (e.g.\ in the mm) would enable the central location of the disc to be determined with a high level of accuracy. 

\subsection{Intensity profiles in the visibility plane}

\begin{figure}
\includegraphics[width=\columnwidth]{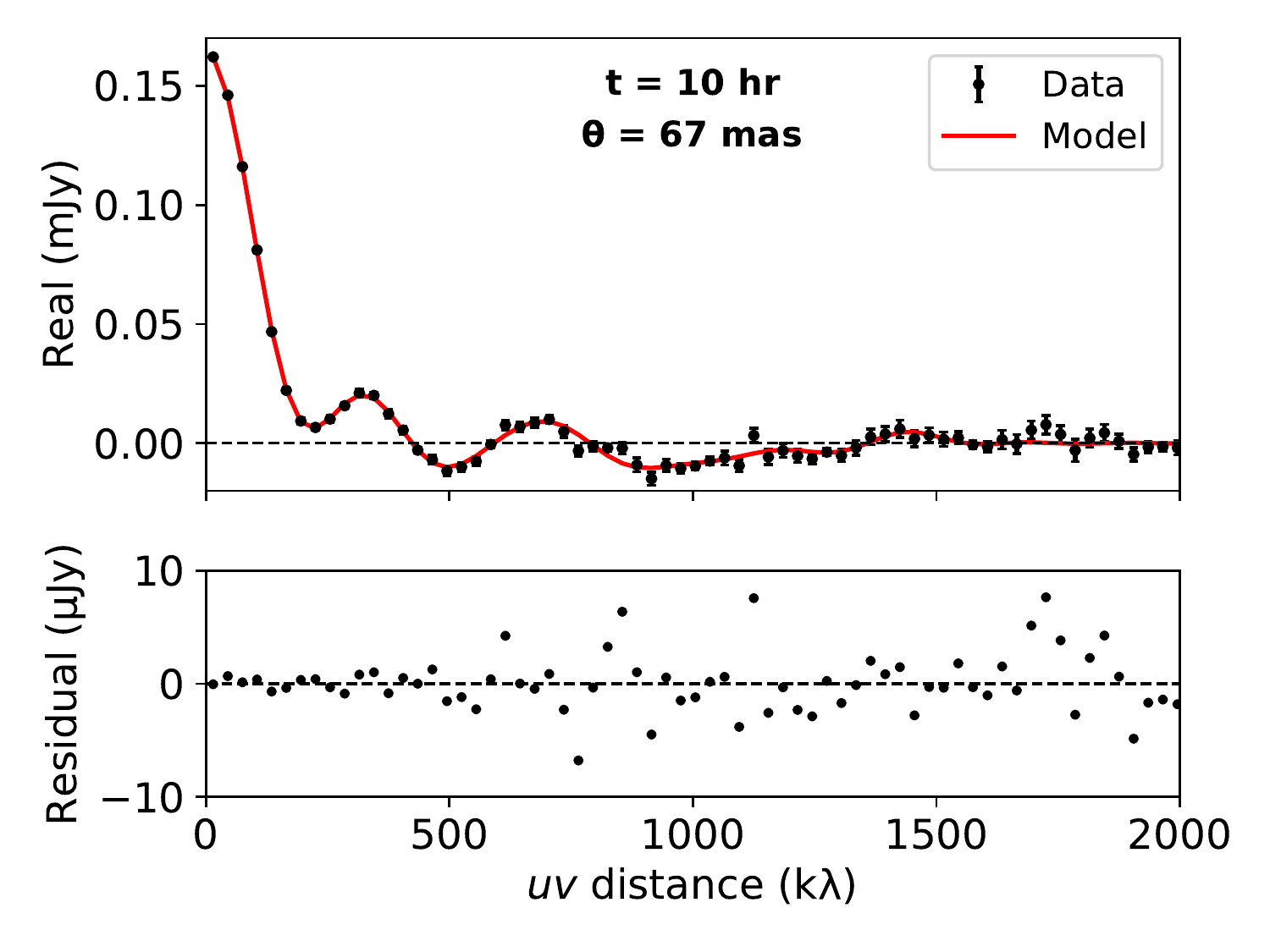}
\caption{Real component of the visibilities (top, black) for the observation of the $a = 1$\,cm model for a 10\,hr integration at 67\,mas binned to a resolution of 30\,k$\lambda$ and truncated at 2000\,k$\lambda$.  The best fitting model is overlaid (top, red) along with the residuals (bottom).}
\label{fig:uv_fit}
\end{figure}

Interferometric image reconstruction relies on estimating the real on-sky brightness distribution from the directly observed visibilities.  However, a number of caveats exist when performing this process -- the visibilities are only finitely sampled, deconvolution algorithms with non-linear effects need to be applied (e.g. CLEAN, \citealt{hogbom_1974}), and image pixels will suffer from the effects of poorly-constrained correlated noise.  The combination of these effects means that images produced in this way can contain a variety of artefacts and (as we have demonstrated in the previous section) require long integration times in order to obtain high fidelity. It is therefore preferable to extract information directly from the visibility plane, and we investigate such an approach on our synthetic observations here, focusing on the $\theta = 67$\,mas case with an integration time of 10\,hrs.

\smallskip

We follow the method of \citet{zhang_2016} and assume that the disc intensity distribution can be treated as a summation over infinitesimally narrow rings, which when Fourier transformed into visibility space can be represented by a zeroth-order Bessel function of the first kind \citep[e.g. $J_0$,][]{berger_2007}.  In this case, the real components are given as  
\begin{equation}
V_\mathrm{Re}(\rho) = 2\pi \int_0^{\infty} I(r) \, J_0(2\pi\,r\,\rho) \, r \, dr,
\label{eqn:equation1}
\end{equation}
where $\rho$ is the baseline grid (for which the binned data grid was used).  Any azimuthally-symmetric intensity distribution, $I(r)$, can be inserted into this equation to compute the corresponding real components of the visibilities.  

\smallskip

To extract the intensity profiles from the binned and deprojected visibilities, we model the intensity distribution as a summation over Gaussians each with amplitude $a_i$ and width $\sigma_i$.  These are also modulated by a sinusoidal function with a spatial frequency $\rho_i$ \citep[see][for further details]{zhang_2016}.  The underlying intensity distribution is also assumed to be a Gaussian function with width $\sigma_0$.  This technique allows the addition of radial substructure (including peaks and troughs) with various spatial frequencies, and ensures that $I \to 0$ as $r \to \infty$.  The total radial intensity profile is described by
\begin{multline}
I_{n}(r) = \frac{a_0}{\sqrt{2\pi}\sigma_0} \exp{\left(-\frac{r^2}{2\sigma_0^2}\right)} \, + \\ \, 
 \sum_{i=1}^{n}\cos{(2\pi r \rho_i)} \times  \frac{a_i}{\sqrt{2\pi}\sigma_i} \exp{\left(-\frac{r^2}{2\sigma_i^2}\right)},   
\end{multline}
where $n$ corresponds to the number Gaussian components.   We use a Markov Chain Monte Carlo approach coupled with Bayesian statistics \citep{patil_2010} to determine the best-fit intensity profile, $I_n(r)$.  A radial grid from 0--150\,au in steps of 1\,au was of sufficient resolution to reproduce the intensity profile. We assume all prior distributions for the fitted parameters are uniform with boundaries set by the range of radii and spatial frequencies within the data (i.e. $-10 < a_{i} < 10$; $0 < \sigma_{i} < 150$\,au; $0 < \rho_{i} < 7000$\,$k\lambda$).  The model intensity profile is then scaled to the observations using the observed flux, which is assumed to have a normal distribution with a standard deviation of 10 per cent at 12.5\,GHz.

\smallskip

We adopt the following procedure -- firstly, a single Gaussian intensity profile is fitted ($I_0$) to determine the best-fit width $\sigma_0$, and to test whether this is an adequate fit to the data.  Additional single Gaussian components are then systematically added to the fit, using the best-fit parameters from the previous model as a starting position.  Further addition of components is stopped when only incremental improvements are found to the residuals of the real component of the visibilities \citep[see][]{walsh_2016}.

\smallskip

We find that four Gaussian components (i.e., $I_4$) are required to adequately reproduce the structure seen in the real component of the visibilities.  This model shows excellent agreement with the observed profile (see Figure \ref{fig:uv_fit}), yielding a reduced $\chi^{2}$ value of 1.49 when compared with the data.  Peak residuals from the best fitting model are of the order 10\,$\mu$Jy and are primarily found at larger $uv$ distances.  The semi-regular oscillation of these residuals may indicate that the addition of more components (reproducing intensity variations at smaller spatial scales) could further improve the fit.  However, as our goal here is to recover the main features of the global structure of the radial intensity profile, and these residuals correspond to less than 5 per cent of the peak flux density for these observations, we do not fit further components. 

\smallskip

\begin{figure}
\includegraphics[width=\columnwidth]{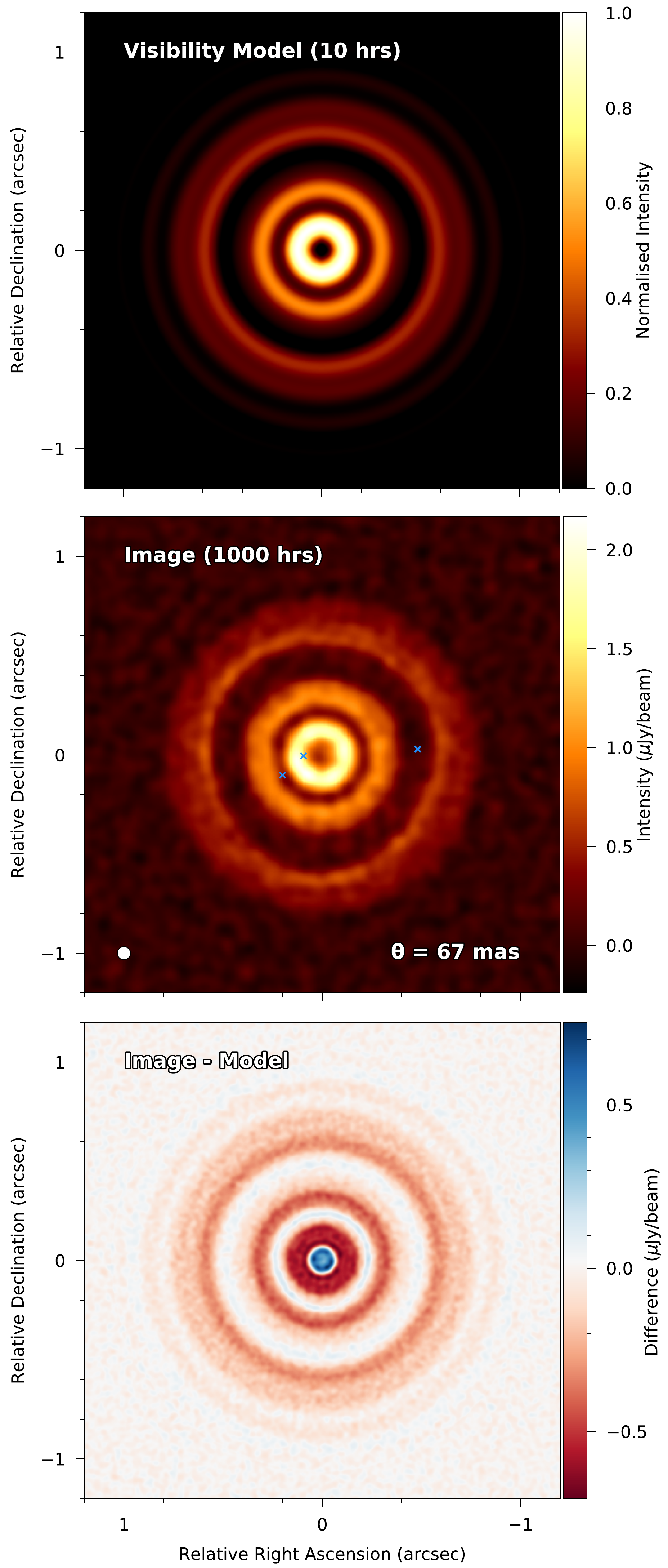}
\caption{Best-fitting model image intensity obtained from the visibilities of the 10\,hr observation at 67 mas (top, see Figure \ref{fig:uv_fit}), compared with the image obtained using an integration time of 1000\,hrs (middle), and the difference between the two (bottom). Modelling the visibilities for the 10\,hr integration reproduces the morphology of the disc extremely well, and recovers features only observed in the image plane using much longer integration times.}
\label{fig:uv_image}
\end{figure}

With the best fitting model for the visibilities available, we can now take the Fourier transform and compare this to observations in the image plane.  Figure \ref{fig:uv_image} shows the resulting model image intensity obtained from fitting the visibilities of the 10\,hr observation (top) alongside the image obtained for an integration time of 1000\,hrs at $\theta=67$\,mas (middle), and finally the difference between these two images (bottom).   It is clear that the model image intensity reproduces very well many of the features of the 1000\,hr integration image (which are not apparent in the 10\,hr integration image).  These include the three bright ring components, the contrast ratios between the rings and gaps, and even the `pile-up' of grains seen toward the inner edge of the outermost ring at $\sim$80\,au ($0\farcs6$).  Examining the difference image, it can be seen that in general the model over-predicts the intensity of the bright rings, but the magnitude of this is similar to the rms in the 10\,hr image, $\sim$0.5\,$\mu$Jy/beam.  There are also differences in the innermost regions of the disc, with the model under-predicting emission within the central $0\farcs1$ (14\,au).  The addition of further components to the best fitting model as discussed above, particularly those to reproduce visibility features at larger $uv$ distances ($\gtrsim 1000$\,k$\lambda$) would improve this discrepancy (but again note that the magnitude is of the order of the 10\,hr rms noise, $\sim$0.5--0.6\,$\mu$Jy/beam).

\smallskip

Similarly to producing a radial profile in the image plane, the success of visibility fitting method we describe here depends strongly on how well the central location of the disc is known.  As discussed previously, this could be accomplished by minimisation of the imaginary component of the visibilities during the fitting procedure.

\section{Discussion}
\label{sec:discussion}

\subsection{Synergies between mm and cm observations}

There is increasing evidence that observations of discs at (sub-)mm wavelengths may not be sensitive probes of their bulk properties.  For example, \citet{carrasco-gonzalez_2019} recently determined that the continuum emission from HL~Tau is optically thick ($\tau = 1$--10) within 60\,au for wavelengths between 0.9--2.1\,mm.  More generally, \citet{zhu_2019} showed that the effects of scattering can lead to optically thick discs being mis-identified as being optically thin at mm wavelengths. Observations of rare CO isotopologues toward protoplanetary discs have also demonstrated that (sub-)mm molecular line emission from discs can be optically thick \citep[e.g. for HD~163296 and HL~Tau,][]{booth_2019, booth_2020}. All of these effects will lead to underestimates of fundamental disc properties, such as their dust and gas masses.  Observations of dust continuum emission at longer wavelengths, such as those available with the SKA, are much more likely to be optically thin.  These will therefore probe the bulk of the disc material even in the most dense regions, providing the most reliable measurements of disc properties.

\smallskip

The combination of observations at multiple frequencies can have also significant advantages.  Measurement of the spectral index of the dust opacity $\beta$ in protoplanetary discs is a common tool in characterising the nature of their dust content, particularly the level of dust growth \citep[see][for a review]{testi_2014}. Accurate recover of $\beta$ requires high precision photometry that, crucially, covers a large enough lever-arm in frequency.  Current studies combining observations with ALMA (at 100's of GHz) and the JVLA (at 30--50\,GHz) have been able to study radial variations in the dust content of many discs \citep[e.g.][]{perez_2015}, confirming predicted phenomena such as radial drift. The recent discovery of small-scale (a few au) substructure in sub-mm observations of protoplanetary disks \citep[e.g.][]{isella_2018, isella_2019} has resulted in numerous theories about their origin (e.g. vortices, instabilities, dust traps, or circumplanetary discs).  Measuring the spectral index, and therefore the properties of the dust grains within these features, is an essential requirement in determining their nature.  However, it is currently not possible to produce maps of azimuthal variations in spectral index across discs at $\sim$au scales due to the limited spatial resolution of observations at lower frequencies.  The joint operation of ALMA and the SKA would enable observations of comparable angular resolution ($\lesssim$40\,mas) at both 100's and 10's of GHz, allowing a detailed characterisation of disc structures down to a few au in the nearest star forming regions.  

\subsection{Disc orientation}

We have assumed that our representative disc model lies at an face-on inclination.  While some discs share a similar orientation (e.g.\ TW Hya, \citealt{andrews_2016}), obviously the majority of discs are observed at a number of intermediate inclinations.  For discs at larger inclinations ($i \gtrsim 60\degr$), characterising substructure becomes more difficult but is still possible \citep[e.g. HD~142666, DoAr~25 \& MY~Lup;][]{huang_2018}. In fact, discs at low-to-intermediate inclinations offer the ability to directly measure the thickness of the emitting layer in the disc vertical direction \citep[as already done for mm-sized grains, see e.g.][]{pinte_2016, villenave_2020}.  Such measurements are able to place strong constraints on the level of vertical mixing in discs due to, for example, turbulence.  Measurement of the thickness of the cm-continuum emitting layer in discs would provide direct observational constraints on the degree of settling of cm-sized particles, the degree of which is important to assess where vertically the cm-grains are located. 

\subsection{Other contributions to cm continuum emission}

Many young star-disc systems exhibit varying degrees of contribution to their cm continuum via sources other than thermal emission from dust grains \citep[see][for a review]{anglada_2018}.  These contributions can take the form of thermal free-free (or bremsstrahlung) emission from ionised jets and winds \citep[e.g.][]{carrasco-gonzalez_2012}, in addition to non-thermal radio gyrosynchrotron emission produced by electrons accelerated in the stellar magnetosphere \citep[e.g.][]{rivilla_2015}.  In order to derive meaningful information on the dust emission, it is necessary to carefully characterise the relative contributions of these possible sources of emission.  In practice, this can be prone to major uncertainties due to small ($\sim$au) size of the emission compared with currently available angular resolutions at these frequencies.  While observations of protoplanetary discs with the SKA will be subject to these effects, the significant increase in spatial resolution will enable a direct observation of the extended emission from the outer disc without contamination (and allow better characterisation of the relative contributions to the continuum emission in the inner disc). Long-term monitoring campaigns will also help to identify the nature of any non-thermal emission, which can be variable in time \citep[e.g.][]{forbrich_2017}.  In addition, the high dynamic range anticipated to be available with SKA continuum observations ($\gtrsim 10^{5}$) will enable characterisation of dust disc emission even in sources with particularly bright jets or flares \citep[see][]{coutens_2019}.

\subsection{Possible sources of interference}

We note that the observing frequencies we utilise in this will be negatively affected by the growth of a number of satellite `mega-constellations', many of which operate between frequencies of 10.7--12.7\,GHz.  Fortunately, this can be mitigated against by splitting the continuum bandwidth between the upper and lower frequencies available within Band 5b (with a possible reduction in sensitivity if a decrease in total bandwidth is required).  Such an observing strategy would also enable measurements of the spectral index across the frequency range offered by Band 5 in a single observation. 

\subsection{A `Young Cluster Deep Field' SKA Key Science Project}

Our results suggest that in order to obtain high fidelity images of cm emission from protoplanetary discs with the SKA, long ($\sim$100--1000\,hr) integration times are required.  While these would be unfeasible in a single pointing, they would be more easily achieved over a longer-term monitoring program.  Such an idea underpins the proposed `Young Cluster Deep Field' SKA Key Science Project \citep[see][]{hoare_2015}, which would utilise the large field of view of the SKA (6.7\arcmin\ at 12.5\,GHz, \citealt{braun_2019}) in order to perform repeated, regular observations of a cluster of young stellar objects.  Such a project has the potential to probe many aspects of star (and planet) formation simultaneously.  One proposed target cluster is the $\rho$ Oph A, which contains 18 sources detected in the cm ranging from Class 0 -- III \citep{coutens_2019}, allowing a range of stellar masses and evolutionary stages to be observed together.  In addition to high SNR observations of the continuum emission, long integrations have also the potential to reveal the complex molecular content in protoplanetary discs \citep[see][]{codella_2015}.  With the SKA design baseline offering 65k channels across 5\,GHz of bandwidth, this would give a channel spacing of 1-3\,km\,s$^{-1}$ across Band 5, allowing line detection studies to be performed simultaneously with continuum observations.  Careful selection of the cadence of the repeated observations would also enable many other science goals to be addressed, including time monitoring of ionised gas in both free-free emission and the emission from hydrogen radio recombination lines (RRLs).  For thermal emission, this would enable characterisation of changes in jet morphology and kinematics (with important links to accretion \&  ejection processes, e.g.\ \citealt{purser_2018, jimenez-serra_2020}).  In addition, ionised gas contributions to the continuum will be able to be quantified utilizing detections of RRLs which scale directly with the ionised continuum emission. For non-thermal emission, this would enable studies of (proto)stellar magnetic flaring activity (which will influence the structure and chemical composition of the the innermost regions and upper layers of protoplanetary discs, e.g.\ \citealt{rab_2017}), along with synchrotron emission from jets. Combining long term monitoring observations with VLBI would also enable mapping of the 6D tomography via non-thermal emission, allowing characterisation of the structure and kinematics of the young cluster itself \citep[see][]{loinard_2014}.  

\section{Conclusions}
\label{sec:conclusions}

This paper presents the first investigation into the observability of dust substructure in protoplanetary discs with the Square Kilometre Array (SKA).  Using a combination of dusty hydrodynamic models and continuum radiative transfer, we examined the capability of the SKA to observe the morphology of cm-sized dust grains (or pebbles) in a disc hosting three forming planets.  We summarise our main findings as follows:  

\begin{itemize}

\item While the planets in our simulated disc are not massive enough to open gaps in the gas, they cause significant annular substructure to be formed in the cm-sized dust within the disc.  These include deep and wide gaps at the radii of the planets themselves, and a `pile-up' of dust at the gap outer edges.  These locations have recently been suggested to be a possible location for `secondary' planet formation in discs \citep[e.g.][]{carrera_2020}, and SKA observations will be key in providing observational constraints for these theories.  

\item Our simulated SKA observations at 12.5\,GHz (the centre of the proposed Band 5b) show that an intermediate angular resolution of 67\,mas provides an excellent trade-off with sensitivity to probe the above disc structures, providing high fidelity images for integration times of 100--1000\,hrs.  Observations at higher angular resolutions (34\,mas) do not result in the recovery of any further substructure, and are limited by SNR even for the longest integration times we study here (1000\,hrs).

\item Characterising the bulk radial structure of the disc will be possible with shorter integration times (tens of hrs) by making use of techniques to improve SNR such as radial averaging in the image plane or directly modelling deprojected visibilities in the $uv$ plane.

\item By conducting our analysis of the intensity distribution in the visibility plane, we show that it is possible to recover disc substructure that would otherwise only be detectable in the image plane using integration times that are 1-2 orders of magnitude longer.   
  
\end{itemize}

Our results demonstrate that SKA will play an essential role in characterising the distribution and properties of cm-sized dust in protoplanetary discs.  Utilising the analysis methods we have described here, along with exploiting the extremely large field of view available with the SKA when compared to facilities such as ALMA and the JVLA, will significantly increase the number of sources that can be characterised in a single pointing.  This will open the door to studying a much larger number of protoplanetary discs across the operational lifetime of the SKA, and provide robust observational tests of planet formation.  From a theoretical point of view, we would strongly encourage further studies that link observed morphology of cm grains with fundamental disc and/or planet properties (similarly to recent efforts for the mm) in order to fully exploit the operation of the SKA in the coming decades.  

\smallskip

In subsequent work, we will examine the potential that the SKA has in characterising the (complex) molecular content of protoplanetary discs, and how to best exploit the joint operation of the SKA and ALMA across the (sub-)mm and cm wavelength regimes.

\section*{Acknowledgements}

We are grateful to the referee, Carlos Carrasco-Gonzalez, for a constructive report and to Robert Braun for helpful discussions on the expected performance of the SKA.  J.D.I., C.W. and M.G.H acknowledges support from the Science and Technology Facilities Council of the United Kingdom (STFC) under ST/R000549/1 and ST/T000287/1.  C.H. is a Winton Fellow and this work has been supported by Winton Philanthropies / The David and Claudia Harding Foundation. I.J.-S. has received partial support from the Spanish FEDER (project number ESP2017-86582-C4-1-R), and State Research Agency (AEI) through project numbers PID2019-105552RB-C41 and MDM-2017-0737 Unidad de Excelencia “Mar\'ia de Maeztu” - Centro de Astrobiolog\'ia (CSIC-INTA).

\smallskip

This research used the ALICE2 High Performance Computing Facility at the University of Leicester. This research also used the DiRAC Data Intensive service at Leicester, operated by the University of Leicester IT Services, which forms part of the STFC DiRAC HPC Facility (\url{www.dirac.ac.uk}). The equipment was funded by BEIS capital funding via STFC capital grants ST/K000373/1 and ST/R002363/1 and STFC DiRAC Operations grant ST/R001014/1. DiRAC is part of the National e-Infrastructure.  We would like to thank Daniel Price for his publicly available SPH plotting code SPLASH \citep{price_2007}, which we have made use of in this paper.

\section*{Data Availability}

The data underlying this article will be shared on reasonable request to the corresponding author.

\bibliographystyle{mnras}

\begin{thebibliography}{}
\makeatletter
\relax
\def\mn@urlcharsother{\let\do\@makeother \do\$\do\&\do\#\do\^\do\_\do\%\do\~}
\def\mn@doi{\begingroup\mn@urlcharsother \@ifnextchar [ {\mn@doi@}
  {\mn@doi@[]}}
\def\mn@doi@[#1]#2{\def\@tempa{#1}\ifx\@tempa\@empty \href
  {http://dx.doi.org/#2} {doi:#2}\else \href {http://dx.doi.org/#2} {#1}\fi
  \endgroup}
\def\mn@eprint#1#2{\mn@eprint@#1:#2::\@nil}
\def\mn@eprint@arXiv#1{\href {http://arxiv.org/abs/#1} {{\tt arXiv:#1}}}
\def\mn@eprint@dblp#1{\href {http://dblp.uni-trier.de/rec/bibtex/#1.xml}
  {dblp:#1}}
\def\mn@eprint@#1:#2:#3:#4\@nil{\def\@tempa {#1}\def\@tempb {#2}\def\@tempc
  {#3}\ifx \@tempc \@empty \let \@tempc \@tempb \let \@tempb \@tempa \fi \ifx
  \@tempb \@empty \def\@tempb {arXiv}\fi \@ifundefined
  {mn@eprint@\@tempb}{\@tempb:\@tempc}{\expandafter \expandafter \csname
  mn@eprint@\@tempb\endcsname \expandafter{\@tempc}}}

\bibitem[\protect\citeauthoryear{{ALMA Partnership} et~al.,}{{ALMA Partnership}
  et~al.}{2015}]{almapartnership2015}
{ALMA Partnership} et~al., 2015, \mn@doi [\apjl] {10.1088/2041-8205/808/1/L3},
  \href {https://ui.adsabs.harvard.edu/abs/2015ApJ...808L...3A} {808, L3}

\bibitem[\protect\citeauthoryear{{Andrews} et~al.,}{{Andrews}
  et~al.}{2012}]{Andrews2012}
{Andrews} S.~M.,  et~al., 2012, \mn@doi [\apj] {10.1088/0004-637X/744/2/162},
  \href {https://ui.adsabs.harvard.edu/abs/2012ApJ...744..162A} {744, 162}

\bibitem[\protect\citeauthoryear{{Andrews} et~al.,}{{Andrews}
  et~al.}{2016}]{andrews_2016}
{Andrews} S.~M.,  et~al., 2016, \mn@doi [\apjl] {10.3847/2041-8205/820/2/L40},
  \href {https://ui.adsabs.harvard.edu/abs/2016ApJ...820L..40A} {820, L40}

\bibitem[\protect\citeauthoryear{{Anglada}, {Rodr{\'\i}guez}  \&
  {Carrasco-Gonz{\'a}lez}}{{Anglada} et~al.}{2018}]{anglada_2018}
{Anglada} G.,  {Rodr{\'\i}guez} L.~F.,   {Carrasco-Gonz{\'a}lez} C.,  2018,
  \mn@doi [\aapr] {10.1007/s00159-018-0107-z}, \href
  {https://ui.adsabs.harvard.edu/abs/2018A&ARv..26....3A} {26, 3}

\bibitem[\protect\citeauthoryear{{Ansdell} et~al.,}{{Ansdell}
  et~al.}{2016}]{ansdelletal2016}
{Ansdell} M.,  et~al., 2016, \mn@doi [\apj] {10.3847/0004-637X/828/1/46}, \href
  {https://ui.adsabs.harvard.edu/abs/2016ApJ...828...46A} {828, 46}

\bibitem[\protect\citeauthoryear{{Bate}, {Bonnell}  \& {Price}}{{Bate}
  et~al.}{1995}]{bate1995}
{Bate} M.~R.,  {Bonnell} I.~A.,   {Price} N.~M.,  1995, \mn@doi [\mnras]
  {10.1093/mnras/277.2.362}, \href
  {https://ui.adsabs.harvard.edu/abs/1995MNRAS.277..362B} {277, 362}

\bibitem[\protect\citeauthoryear{{Berger} \& {Segransan}}{{Berger} \&
  {Segransan}}{2007}]{berger_2007}
{Berger} J.~P.,  {Segransan} D.,  2007, \mn@doi [\nar]
  {10.1016/j.newar.2007.06.003}, \href
  {https://ui.adsabs.harvard.edu/abs/2007NewAR..51..576B} {51, 576}

\bibitem[\protect\citeauthoryear{{Birnstiel} et~al.,}{{Birnstiel}
  et~al.}{2018}]{birnstiel_2018}
{Birnstiel} T.,  et~al., 2018, \mn@doi [\apjl] {10.3847/2041-8213/aaf743},
  \href {https://ui.adsabs.harvard.edu/abs/2018ApJ...869L..45B} {869, L45}

\bibitem[\protect\citeauthoryear{{Blum} \& {Wurm}}{{Blum} \&
  {Wurm}}{2008}]{blum_2008}
{Blum} J.,  {Wurm} G.,  2008, \mn@doi [\araa]
  {10.1146/annurev.astro.46.060407.145152}, \href
  {https://ui.adsabs.harvard.edu/abs/2008ARA&A..46...21B} {46, 21}

\bibitem[\protect\citeauthoryear{{Booth} \& {Ilee}}{{Booth} \&
  {Ilee}}{2020}]{booth_2020}
{Booth} A.~S.,  {Ilee} J.~D.,  2020, \mn@doi [\mnras] {10.1093/mnrasl/slaa014},
  \href {https://ui.adsabs.harvard.edu/abs/2020MNRAS.493L.108B} {493, L108}

\bibitem[\protect\citeauthoryear{Booth, Walsh, Ilee, Notsu, Qi, Nomura  \&
  Akiyama}{Booth et~al.}{2019}]{booth_2019}
Booth A.~S.,  Walsh C.,  Ilee J.~D.,  Notsu S.,  Qi C.,  Nomura H.,   Akiyama
  E.,  2019, \mn@doi [The Astrophysical Journal] {10.3847/2041-8213/ab3645},
  882, L31

\bibitem[\protect\citeauthoryear{{Braun}, {Bourke}, {Green}, {Keane}  \&
  {Wagg}}{{Braun} et~al.}{2015}]{braun_2015}
{Braun} R.,  {Bourke} T.,  {Green} J.~A.,  {Keane} E.,   {Wagg} J.,  2015, in
  Advancing Astrophysics with the Square Kilometre Array (AASKA14). p.~174

\bibitem[\protect\citeauthoryear{{Braun}, {Bonaldi}, {Bourke}, {Keane}  \&
  {Wagg}}{{Braun} et~al.}{2019}]{braun_2019}
{Braun} R.,  {Bonaldi} A.,  {Bourke} T.,  {Keane} E.,   {Wagg} J.,  2019, arXiv
  e-prints, \href {https://ui.adsabs.harvard.edu/abs/2019arXiv191212699B} {p.
  arXiv:1912.12699}

\bibitem[\protect\citeauthoryear{{Carrasco-Gonz{\'a}lez}
  et~al.,}{{Carrasco-Gonz{\'a}lez} et~al.}{2012}]{carrasco-gonzalez_2012}
{Carrasco-Gonz{\'a}lez} C.,  et~al., 2012, \mn@doi [\apjl]
  {10.1088/2041-8205/752/2/L29}, \href
  {https://ui.adsabs.harvard.edu/abs/2012ApJ...752L..29C} {752, L29}

\bibitem[\protect\citeauthoryear{{Carrasco-Gonz{\'a}lez}
  et~al.,}{{Carrasco-Gonz{\'a}lez} et~al.}{2019}]{carrasco-gonzalez_2019}
{Carrasco-Gonz{\'a}lez} C.,  et~al., 2019, \mn@doi [\apj]
  {10.3847/1538-4357/ab3d33}, \href
  {https://ui.adsabs.harvard.edu/abs/2019ApJ...883...71C} {883, 71}

\bibitem[\protect\citeauthoryear{{Carrera}, {Simon}, {Li}, {Kretke}  \&
  {Klahr}}{{Carrera} et~al.}{2020}]{carrera_2020}
{Carrera} D.,  {Simon} J.~B.,  {Li} R.,  {Kretke} K.~A.,   {Klahr} H.,  2020,
  arXiv e-prints, \href {https://ui.adsabs.harvard.edu/abs/2020arXiv200801727C}
  {p. arXiv:2008.01727}

\bibitem[\protect\citeauthoryear{{Clarke} et~al.,}{{Clarke}
  et~al.}{2018}]{clarke_2018}
{Clarke} C.~J.,  et~al., 2018, \mn@doi [\apjl] {10.3847/2041-8213/aae36b},
  \href {https://ui.adsabs.harvard.edu/abs/2018ApJ...866L...6C} {866, L6}

\bibitem[\protect\citeauthoryear{{Cleeves}, {{\"O}berg}, {Wilner}, {Huang},
  {Loomis}, {Andrews}  \& {Czekala}}{{Cleeves} et~al.}{2016}]{cleevesetal2016}
{Cleeves} L.~I.,  {{\"O}berg} K.~I.,  {Wilner} D.~J.,  {Huang} J.,  {Loomis}
  R.~A.,  {Andrews} S.~M.,   {Czekala} I.,  2016, \mn@doi [\apj]
  {10.3847/0004-637X/832/2/110}, \href
  {https://ui.adsabs.harvard.edu/abs/2016ApJ...832..110C} {832, 110}

\bibitem[\protect\citeauthoryear{{Codella} et~al.,}{{Codella}
  et~al.}{2015}]{codella_2015}
{Codella} C.,  et~al., 2015, in Advancing Astrophysics with the Square
  Kilometre Array (AASKA14). p.~123 (\mn@eprint {arXiv} {1412.8611})

\bibitem[\protect\citeauthoryear{{Coutens} et~al.,}{{Coutens}
  et~al.}{2019}]{coutens_2019}
{Coutens} A.,  et~al., 2019, \mn@doi [\aap] {10.1051/0004-6361/201935340},
  \href {https://ui.adsabs.harvard.edu/abs/2019A&A...631A..58C} {631, A58}

\bibitem[\protect\citeauthoryear{{Crida}, {Morbidelli}  \& {Masset}}{{Crida}
  et~al.}{2006}]{crida_2006}
{Crida} A.,  {Morbidelli} A.,   {Masset} F.,  2006, \mn@doi [\icarus]
  {10.1016/j.icarus.2005.10.007}, \href
  {https://ui.adsabs.harvard.edu/abs/2006Icar..181..587C} {181, 587}

\bibitem[\protect\citeauthoryear{{Dipierro}, {Price}, {Laibe}, {Hirsh},
  {Cerioli}  \& {Lodato}}{{Dipierro} et~al.}{2015}]{dipierro_2015_hltau}
{Dipierro} G.,  {Price} D.,  {Laibe} G.,  {Hirsh} K.,  {Cerioli} A.,   {Lodato}
  G.,  2015, \mn@doi [\mnras] {10.1093/mnrasl/slv105}, \href
  {https://ui.adsabs.harvard.edu/abs/2015MNRAS.453L..73D} {453, L73}

\bibitem[\protect\citeauthoryear{{Dipierro} et~al.,}{{Dipierro}
  et~al.}{2018}]{dipierroetal2018}
{Dipierro} G.,  et~al., 2018, \mn@doi [\mnras] {10.1093/mnras/sty181}, \href
  {https://ui.adsabs.harvard.edu/abs/2018MNRAS.475.5296D} {475, 5296}

\bibitem[\protect\citeauthoryear{{Dong}, {Zhu}  \& {Whitney}}{{Dong}
  et~al.}{2015}]{Dong2015}
{Dong} R.,  {Zhu} Z.,   {Whitney} B.,  2015, \mn@doi [\apj]
  {10.1088/0004-637X/809/1/93}, \href
  {https://ui.adsabs.harvard.edu/abs/2015ApJ...809...93D} {809, 93}

\bibitem[\protect\citeauthoryear{{Dong}, {Li}, {Chiang}  \& {Li}}{{Dong}
  et~al.}{2018}]{dong_2018}
{Dong} R.,  {Li} S.,  {Chiang} E.,   {Li} H.,  2018, \mn@doi [\apj]
  {10.3847/1538-4357/aadadd}, \href
  {https://ui.adsabs.harvard.edu/abs/2018ApJ...866..110D} {866, 110}

\bibitem[\protect\citeauthoryear{Draine}{Draine}{2006}]{draine_2006}
Draine B.~T.,  2006, \mn@doi [The Astrophysical Journal] {10.1086/498130}, 636,
  1114

\bibitem[\protect\citeauthoryear{{Draine} \& {Lee}}{{Draine} \&
  {Lee}}{1984}]{drainelee1984}
{Draine} B.~T.,  {Lee} H.~M.,  1984, \mn@doi [\apj] {10.1086/162480}, \href
  {https://ui.adsabs.harvard.edu/abs/1984ApJ...285...89D} {285, 89}

\bibitem[\protect\citeauthoryear{{Dullemond} et~al.,}{{Dullemond}
  et~al.}{2018}]{dullemond_2018}
{Dullemond} C.~P.,  et~al., 2018, \mn@doi [\apjl] {10.3847/2041-8213/aaf742},
  \href {https://ui.adsabs.harvard.edu/abs/2018ApJ...869L..46D} {869, L46}

\bibitem[\protect\citeauthoryear{{Fedele} et~al.,}{{Fedele}
  et~al.}{2017}]{fedele_2017}
{Fedele} D.,  et~al., 2017, \mn@doi [\aap] {10.1051/0004-6361/201629860}, \href
  {https://ui.adsabs.harvard.edu/abs/2017A&A...600A..72F} {600, A72}

\bibitem[\protect\citeauthoryear{{Fedele} et~al.,}{{Fedele}
  et~al.}{2018}]{fedele_2018}
{Fedele} D.,  et~al., 2018, \mn@doi [\aap] {10.1051/0004-6361/201731978}, \href
  {https://ui.adsabs.harvard.edu/abs/2018A&A...610A..24F} {610, A24}

\bibitem[\protect\citeauthoryear{{Flock}, {Ruge}, {Dzyurkevich}, {Henning},
  {Klahr}  \& {Wolf}}{{Flock} et~al.}{2015}]{Flock2015}
{Flock} M.,  {Ruge} J.~P.,  {Dzyurkevich} N.,  {Henning} T.,  {Klahr} H.,
  {Wolf} S.,  2015, \mn@doi [\aap] {10.1051/0004-6361/201424693}, \href
  {https://ui.adsabs.harvard.edu/abs/2015A&A...574A..68F} {574, A68}

\bibitem[\protect\citeauthoryear{{Forbrich}, {Reid}, {Menten}, {Rivilla},
  {Wolk}, {Rau}  \& {Chand ler}}{{Forbrich} et~al.}{2017}]{forbrich_2017}
{Forbrich} J.,  {Reid} M.~J.,  {Menten} K.~M.,  {Rivilla} V.~M.,  {Wolk} S.~J.,
   {Rau} U.,   {Chand ler} C.~J.,  2017, \mn@doi [\apj]
  {10.3847/1538-4357/aa7aa4}, \href
  {https://ui.adsabs.harvard.edu/abs/2017ApJ...844..109F} {844, 109}

\bibitem[\protect\citeauthoryear{{Guidi} et~al.,}{{Guidi}
  et~al.}{2016}]{guidi_2016}
{Guidi} G.,  et~al., 2016, \mn@doi [\aap] {10.1051/0004-6361/201527516}, \href
  {https://ui.adsabs.harvard.edu/abs/2016A&A...588A.112G} {588, A112}

\bibitem[\protect\citeauthoryear{{Hoare} et~al.,}{{Hoare}
  et~al.}{2015}]{hoare_2015}
{Hoare} M.,  et~al., 2015, in Advancing Astrophysics with the Square Kilometre
  Array (AASKA14). p.~115

\bibitem[\protect\citeauthoryear{{H{\"o}gbom}}{{H{\"o}gbom}}{1974}]{hogbom_1974}
{H{\"o}gbom} J.~A.,  1974, \aaps, \href
  {https://ui.adsabs.harvard.edu/abs/1974A&AS...15..417H} {15, 417}

\bibitem[\protect\citeauthoryear{{Huang} et~al.,}{{Huang}
  et~al.}{2018}]{huang_2018}
{Huang} J.,  et~al., 2018, \mn@doi [\apjl] {10.3847/2041-8213/aaf740}, \href
  {https://ui.adsabs.harvard.edu/abs/2018ApJ...869L..42H} {869, L42}

\bibitem[\protect\citeauthoryear{{Isella} et~al.,}{{Isella}
  et~al.}{2018}]{isella_2018}
{Isella} A.,  et~al., 2018, \mn@doi [\apjl] {10.3847/2041-8213/aaf747}, \href
  {https://ui.adsabs.harvard.edu/abs/2018ApJ...869L..49I} {869, L49}

\bibitem[\protect\citeauthoryear{{Isella}, {Benisty}, {Teague}, {Bae},
  {Keppler}, {Facchini}  \& {P{\'e}rez}}{{Isella} et~al.}{2019}]{isella_2019}
{Isella} A.,  {Benisty} M.,  {Teague} R.,  {Bae} J.,  {Keppler} M.,  {Facchini}
  S.,   {P{\'e}rez} L.,  2019, \mn@doi [\apjl] {10.3847/2041-8213/ab2a12},
  \href {https://ui.adsabs.harvard.edu/abs/2019ApJ...879L..25I} {879, L25}

\bibitem[\protect\citeauthoryear{{Jim{\'e}nez-Serra}, {B{\'a}ez-Rubio},
  {Mart{\'\i}n-Pintado}, {Zhang}  \& {Rivilla}}{{Jim{\'e}nez-Serra}
  et~al.}{2020}]{jimenez-serra_2020}
{Jim{\'e}nez-Serra} I.,  {B{\'a}ez-Rubio} A.,  {Mart{\'\i}n-Pintado} J.,
  {Zhang} Q.,   {Rivilla} V.~M.,  2020, \mn@doi [\apjl]
  {10.3847/2041-8213/aba050}, \href
  {https://ui.adsabs.harvard.edu/abs/2020ApJ...897L..33J} {897, L33}

\bibitem[\protect\citeauthoryear{{Johansen}, {Oishi}, {Mac Low}, {Klahr},
  {Henning}  \& {Youdin}}{{Johansen} et~al.}{2007}]{johansen_2007}
{Johansen} A.,  {Oishi} J.~S.,  {Mac Low} M.-M.,  {Klahr} H.,  {Henning} T.,
  {Youdin} A.,  2007, \mn@doi [\nat] {10.1038/nature06086}, \href
  {https://ui.adsabs.harvard.edu/abs/2007Natur.448.1022J} {448, 1022}

\bibitem[\protect\citeauthoryear{{Kanagawa}, {Muto}, {Tanaka}, {Tanigawa},
  {Takeuchi}, {Tsukagoshi}  \& {Momose}}{{Kanagawa}
  et~al.}{2015}]{kanagawa_2015}
{Kanagawa} K.~D.,  {Muto} T.,  {Tanaka} H.,  {Tanigawa} T.,  {Takeuchi} T.,
  {Tsukagoshi} T.,   {Momose} M.,  2015, \mn@doi [\apjl]
  {10.1088/2041-8205/806/1/L15}, \href
  {https://ui.adsabs.harvard.edu/abs/2015ApJ...806L..15K} {806, L15}

\bibitem[\protect\citeauthoryear{{Keppler} et~al.,}{{Keppler}
  et~al.}{2018}]{keppler_2018}
{Keppler} M.,  et~al., 2018, \mn@doi [\aap] {10.1051/0004-6361/201832957},
  \href {https://ui.adsabs.harvard.edu/abs/2018A&A...617A..44K} {617, A44}

\bibitem[\protect\citeauthoryear{{Lagrange} et~al.,}{{Lagrange}
  et~al.}{2010}]{lagrange_2010}
{Lagrange} A.~M.,  et~al., 2010, \mn@doi [Science] {10.1126/science.1187187},
  \href {https://ui.adsabs.harvard.edu/abs/2010Sci...329...57L} {329, 57}

\bibitem[\protect\citeauthoryear{{Laibe} \& {Price}}{{Laibe} \&
  {Price}}{2012a}]{laibeprice2012a}
{Laibe} G.,  {Price} D.~J.,  2012a, \mn@doi [\mnras]
  {10.1111/j.1365-2966.2011.20202.x}, \href
  {https://ui.adsabs.harvard.edu/abs/2012MNRAS.420.2345L} {420, 2345}

\bibitem[\protect\citeauthoryear{{Laibe} \& {Price}}{{Laibe} \&
  {Price}}{2012b}]{laibeprice2012b}
{Laibe} G.,  {Price} D.~J.,  2012b, \mn@doi [\mnras]
  {10.1111/j.1365-2966.2011.20201.x}, \href
  {https://ui.adsabs.harvard.edu/abs/2012MNRAS.420.2365L} {420, 2365}

\bibitem[\protect\citeauthoryear{{Lebreuilly}, {Commer{\c{c}}on}  \&
  {Laibe}}{{Lebreuilly} et~al.}{2020}]{lebreuilly_2020}
{Lebreuilly} U.,  {Commer{\c{c}}on} B.,   {Laibe} G.,  2020, arXiv e-prints,
  \href {https://ui.adsabs.harvard.edu/abs/2020arXiv200706050L} {p.
  arXiv:2007.06050}

\bibitem[\protect\citeauthoryear{{Loinard} et~al.,}{{Loinard}
  et~al.}{2014}]{loinard_2014}
{Loinard} L.,  et~al., 2014, arXiv e-prints, \href
  {https://ui.adsabs.harvard.edu/abs/2014arXiv1412.6481L} {p. arXiv:1412.6481}

\bibitem[\protect\citeauthoryear{{Long} et~al.,}{{Long}
  et~al.}{2019}]{Long2019}
{Long} F.,  et~al., 2019, \mn@doi [\apj] {10.3847/1538-4357/ab2d2d}, \href
  {https://ui.adsabs.harvard.edu/abs/2019ApJ...882...49L} {882, 49}

\bibitem[\protect\citeauthoryear{{Marois}, {Macintosh}, {Barman}, {Zuckerman},
  {Song}, {Patience}, {Lafreni{\`e}re}  \& {Doyon}}{{Marois}
  et~al.}{2008}]{marois_2008}
{Marois} C.,  {Macintosh} B.,  {Barman} T.,  {Zuckerman} B.,  {Song} I.,
  {Patience} J.,  {Lafreni{\`e}re} D.,   {Doyon} R.,  2008, \mn@doi [Science]
  {10.1126/science.1166585}, \href
  {https://ui.adsabs.harvard.edu/abs/2008Sci...322.1348M} {322, 1348}

\bibitem[\protect\citeauthoryear{{Mathis}, {Rumpl}  \& {Nordsieck}}{{Mathis}
  et~al.}{1977}]{mathisetal1977}
{Mathis} J.~S.,  {Rumpl} W.,   {Nordsieck} K.~H.,  1977, \mn@doi [\apj]
  {10.1086/155591}, \href
  {https://ui.adsabs.harvard.edu/abs/1977ApJ...217..425M} {217, 425}

\bibitem[\protect\citeauthoryear{{McMullin}, {Waters}, {Schiebel}, {Young}  \&
  {Golap}}{{McMullin} et~al.}{2007}]{mcmullin_2007}
{McMullin} J.~P.,  {Waters} B.,  {Schiebel} D.,  {Young} W.,   {Golap} K.,
  2007, in {Shaw} R.~A.,  {Hill} F.,   {Bell} D.~J.,  eds,  Astronomical
  Society of the Pacific Conference Series Vol. 376, Astronomical Data Analysis
  Software and Systems XVI. p.~127

\bibitem[\protect\citeauthoryear{{Meru}, {Rosotti}, {Booth}, {Nazari}  \&
  {Clarke}}{{Meru} et~al.}{2019}]{meru_2019}
{Meru} F.,  {Rosotti} G.~P.,  {Booth} R.~A.,  {Nazari} P.,   {Clarke} C.~J.,
  2019, \mn@doi [\mnras] {10.1093/mnras/sty2847}, \href
  {https://ui.adsabs.harvard.edu/abs/2019MNRAS.482.3678M} {482, 3678}

\bibitem[\protect\citeauthoryear{{Nayakshin} et~al.,}{{Nayakshin}
  et~al.}{2020}]{nayakshinetal2020}
{Nayakshin} S.,  et~al., 2020, \mn@doi [\mnras] {10.1093/mnras/staa1132}, \href
  {https://ui.adsabs.harvard.edu/abs/2020MNRAS.495..285N} {495, 285}

\bibitem[\protect\citeauthoryear{{Nazari}, {Booth}, {Clarke}, {Rosotti},
  {Tazzari}, {Juhasz}  \& {Meru}}{{Nazari} et~al.}{2019}]{nazari_2019}
{Nazari} P.,  {Booth} R.~A.,  {Clarke} C.~J.,  {Rosotti} G.~P.,  {Tazzari} M.,
  {Juhasz} A.,   {Meru} F.,  2019, \mn@doi [\mnras] {10.1093/mnras/stz836},
  \href {https://ui.adsabs.harvard.edu/abs/2019MNRAS.485.5914N} {485, 5914}

\bibitem[\protect\citeauthoryear{{Okuzumi}, {Momose}, {Sirono}, {Kobayashi}  \&
  {Tanaka}}{{Okuzumi} et~al.}{2016}]{Okuzumi2016}
{Okuzumi} S.,  {Momose} M.,  {Sirono} S.-i.,  {Kobayashi} H.,   {Tanaka} H.,
  2016, \mn@doi [\apj] {10.3847/0004-637X/821/2/82}, \href
  {https://ui.adsabs.harvard.edu/abs/2016ApJ...821...82O} {821, 82}

\bibitem[\protect\citeauthoryear{{Paardekooper} \& {Mellema}}{{Paardekooper} \&
  {Mellema}}{2004}]{Paardekooper2004}
{Paardekooper} S.~J.,  {Mellema} G.,  2004, \mn@doi [\aap]
  {10.1051/0004-6361:200400053}, \href
  {https://ui.adsabs.harvard.edu/abs/2004A&A...425L...9P} {425, L9}

\bibitem[\protect\citeauthoryear{Patil, Huard  \& Fonnesbeck}{Patil
  et~al.}{2010}]{patil_2010}
Patil A.,  Huard D.,   Fonnesbeck C.,  2010, \mn@doi [Journal of Statistical
  Software, Articles] {10.18637/jss.v035.i04}, 35, 1

\bibitem[\protect\citeauthoryear{{P{\'e}rez} et~al.,}{{P{\'e}rez}
  et~al.}{2012}]{perez_2012}
{P{\'e}rez} L.~M.,  et~al., 2012, \mn@doi [\apjl]
  {10.1088/2041-8205/760/1/L17}, \href
  {https://ui.adsabs.harvard.edu/abs/2012ApJ...760L..17P} {760, L17}

\bibitem[\protect\citeauthoryear{{Perez}, {Dunhill}, {Casassus}, {Roman},
  {Szul{\'a}gyi}, {Flores}, {Marino}  \& {Montesinos}}{{Perez}
  et~al.}{2015a}]{perezetal2015}
{Perez} S.,  {Dunhill} A.,  {Casassus} S.,  {Roman} P.,  {Szul{\'a}gyi} J.,
  {Flores} C.,  {Marino} S.,   {Montesinos} M.,  2015a, \mn@doi [\apjl]
  {10.1088/2041-8205/811/1/L5}, \href
  {https://ui.adsabs.harvard.edu/abs/2015ApJ...811L...5P} {811, L5}

\bibitem[\protect\citeauthoryear{{P{\'e}rez} et~al.,}{{P{\'e}rez}
  et~al.}{2015b}]{perez_2015}
{P{\'e}rez} L.~M.,  et~al., 2015b, \mn@doi [\apj] {10.1088/0004-637X/813/1/41},
  \href {https://ui.adsabs.harvard.edu/abs/2015ApJ...813...41P} {813, 41}

\bibitem[\protect\citeauthoryear{{Pinilla}, {de Juan Ovelar}, {Ataiee},
  {Benisty}, {Birnstiel}, {van Dishoeck}  \& {Min}}{{Pinilla}
  et~al.}{2015}]{pinilla_2015}
{Pinilla} P.,  {de Juan Ovelar} M.,  {Ataiee} S.,  {Benisty} M.,  {Birnstiel}
  T.,  {van Dishoeck} E.~F.,   {Min} M.,  2015, \mn@doi [\aap]
  {10.1051/0004-6361/201424679}, \href
  {https://ui.adsabs.harvard.edu/abs/2015A&A...573A...9P} {573, A9}

\bibitem[\protect\citeauthoryear{{Pinte}, {M{\'e}nard}, {Duch{\^e}ne}  \&
  {Bastien}}{{Pinte} et~al.}{2006}]{mcfost}
{Pinte} C.,  {M{\'e}nard} F.,  {Duch{\^e}ne} G.,   {Bastien} P.,  2006, \mn@doi
  [\aap] {10.1051/0004-6361:20053275}, \href
  {https://ui.adsabs.harvard.edu/abs/2006A&A...459..797P} {459, 797}

\bibitem[\protect\citeauthoryear{{Pinte}, {Dent}, {M{\'e}nard}, {Hales},
  {Hill}, {Cortes}  \& {de Gregorio-Monsalvo}}{{Pinte}
  et~al.}{2016}]{pinte_2016}
{Pinte} C.,  {Dent} W.~R.~F.,  {M{\'e}nard} F.,  {Hales} A.,  {Hill} T.,
  {Cortes} P.,   {de Gregorio-Monsalvo} I.,  2016, \mn@doi [\apj]
  {10.3847/0004-637X/816/1/25}, \href
  {https://ui.adsabs.harvard.edu/abs/2016ApJ...816...25P} {816, 25}

\bibitem[\protect\citeauthoryear{{Pinte} et~al.,}{{Pinte}
  et~al.}{2019}]{pinte_2019}
{Pinte} C.,  et~al., 2019, \mn@doi [Nature Astronomy]
  {10.1038/s41550-019-0852-6}, \href
  {https://ui.adsabs.harvard.edu/abs/2019NatAs...3.1109P} {3, 1109}

\bibitem[\protect\citeauthoryear{{Pinte} et~al.,}{{Pinte}
  et~al.}{2020}]{pinte_2020}
{Pinte} C.,  et~al., 2020, \mn@doi [\apjl] {10.3847/2041-8213/ab6dda}, \href
  {https://ui.adsabs.harvard.edu/abs/2020ApJ...890L...9P} {890, L9}

\bibitem[\protect\citeauthoryear{{Price}}{{Price}}{2007}]{price_2007}
{Price} D.~J.,  2007, \mn@doi [\pasa] {10.1071/AS07022}, \href
  {https://ui.adsabs.harvard.edu/abs/2007PASA...24..159P} {24, 159}

\bibitem[\protect\citeauthoryear{{Price} et~al.,}{{Price}
  et~al.}{2018}]{phantom}
{Price} D.~J.,  et~al., 2018, \mn@doi [\pasa] {10.1017/pasa.2018.25}, \href
  {https://ui.adsabs.harvard.edu/abs/2018PASA...35...31P} {35, e031}

\bibitem[\protect\citeauthoryear{{Purser}, {Ainsworth}, {Ray}, {Green},
  {Taylor}  \& {Scaife}}{{Purser} et~al.}{2018}]{purser_2018}
{Purser} S.~J.~D.,  {Ainsworth} R.~E.,  {Ray} T.~P.,  {Green} D.~A.,  {Taylor}
  A.~M.,   {Scaife} A.~M.~M.,  2018, \mn@doi [\mnras] {10.1093/mnras/sty2649},
  \href {https://ui.adsabs.harvard.edu/abs/2018MNRAS.481.5532P} {481, 5532}

\bibitem[\protect\citeauthoryear{{Rab}, {G{\"u}del}, {Padovani}, {Kamp}, {Thi},
  {Woitke}  \& {Aresu}}{{Rab} et~al.}{2017}]{rab_2017}
{Rab} C.,  {G{\"u}del} M.,  {Padovani} M.,  {Kamp} I.,  {Thi} W.~F.,  {Woitke}
  P.,   {Aresu} G.,  2017, \mn@doi [\aap] {10.1051/0004-6361/201630241}, \href
  {https://ui.adsabs.harvard.edu/abs/2017A&A...603A..96R} {603, A96}

\bibitem[\protect\citeauthoryear{{Ribas}, {Mer{\'\i}n}, {Bouy}  \&
  {Maud}}{{Ribas} et~al.}{2014}]{ribas_2014}
{Ribas} {\'A}.,  {Mer{\'\i}n} B.,  {Bouy} H.,   {Maud} L.~T.,  2014, \mn@doi
  [\aap] {10.1051/0004-6361/201322597}, \href
  {https://ui.adsabs.harvard.edu/abs/2014A&A...561A..54R} {561, A54}

\bibitem[\protect\citeauthoryear{{Riols} \& {Lesur}}{{Riols} \&
  {Lesur}}{2019}]{Riols2019}
{Riols} A.,  {Lesur} G.,  2019, \mn@doi [\aap] {10.1051/0004-6361/201834813},
  \href {https://ui.adsabs.harvard.edu/abs/2019A&A...625A.108R} {625, A108}

\bibitem[\protect\citeauthoryear{{Rivilla}, {Chandler}, {Sanz-Forcada},
  {Jim{\'e}nez-Serra}, {Forbrich}  \& {Mart{\'\i}n-Pintado}}{{Rivilla}
  et~al.}{2015}]{rivilla_2015}
{Rivilla} V.~M.,  {Chandler} C.~J.,  {Sanz-Forcada} J.,  {Jim{\'e}nez-Serra}
  I.,  {Forbrich} J.,   {Mart{\'\i}n-Pintado} J.,  2015, \mn@doi [\apj]
  {10.1088/0004-637X/808/2/146}, \href
  {https://ui.adsabs.harvard.edu/abs/2015ApJ...808..146R} {808, 146}

\bibitem[\protect\citeauthoryear{{Rosotti}, {Juhasz}, {Booth}  \&
  {Clarke}}{{Rosotti} et~al.}{2016}]{rosotti_2016}
{Rosotti} G.~P.,  {Juhasz} A.,  {Booth} R.~A.,   {Clarke} C.~J.,  2016, \mn@doi
  [\mnras] {10.1093/mnras/stw691}, \href
  {https://ui.adsabs.harvard.edu/abs/2016MNRAS.459.2790R} {459, 2790}

\bibitem[\protect\citeauthoryear{{Safronov}}{{Safronov}}{1967}]{safronov_1967}
{Safronov} V.~S.,  1967, \sovast, \href
  {https://ui.adsabs.harvard.edu/abs/1967SvA....10..650S} {10, 650}

\bibitem[\protect\citeauthoryear{{Semenov}, {Henning}, {Helling}, {Ilgner}  \&
  {Sedlmayr}}{{Semenov} et~al.}{2003}]{semenov_2003}
{Semenov} D.,  {Henning} T.,  {Helling} C.,  {Ilgner} M.,   {Sedlmayr} E.,
  2003, \mn@doi [\aap] {10.1051/0004-6361:20031279}, \href
  {https://ui.adsabs.harvard.edu/abs/2003A&A...410..611S} {410, 611}

\bibitem[\protect\citeauthoryear{{Shakura} \& {Sunyaev}}{{Shakura} \&
  {Sunyaev}}{1973}]{ss1973}
{Shakura} N.~I.,  {Sunyaev} R.~A.,  1973, \aap, \href
  {https://ui.adsabs.harvard.edu/abs/1973A&A....24..337S} {500, 33}

\bibitem[\protect\citeauthoryear{{Suriano}, {Li}, {Krasnopolsky}  \&
  {Shang}}{{Suriano} et~al.}{2018}]{Suriano2018}
{Suriano} S.~S.,  {Li} Z.-Y.,  {Krasnopolsky} R.,   {Shang} H.,  2018, \mn@doi
  [\mnras] {10.1093/mnras/sty717}, \href
  {https://ui.adsabs.harvard.edu/abs/2018MNRAS.477.1239S} {477, 1239}

\bibitem[\protect\citeauthoryear{{Tapia}, {Lizano}, {Sierra},
  {Carrasco-Gonz{\'a}lez}  \& {Bayona-Bobadilla}}{{Tapia}
  et~al.}{2019}]{tapia_2019}
{Tapia} C.,  {Lizano} S.,  {Sierra} A.,  {Carrasco-Gonz{\'a}lez} C.,
  {Bayona-Bobadilla} E.,  2019, arXiv e-prints, \href
  {https://ui.adsabs.harvard.edu/abs/2019arXiv191105108T} {p. arXiv:1911.05108}

\bibitem[\protect\citeauthoryear{{Tazzari} et~al.,}{{Tazzari}
  et~al.}{2016}]{tazzari_2016}
{Tazzari} M.,  et~al., 2016, \mn@doi [\aap] {10.1051/0004-6361/201527423},
  \href {https://ui.adsabs.harvard.edu/abs/2016A&A...588A..53T} {588, A53}

\bibitem[\protect\citeauthoryear{{Teague}, {Bae}, {Bergin}, {Birnstiel}  \&
  {Foreman-Mackey}}{{Teague} et~al.}{2018}]{teague_2018}
{Teague} R.,  {Bae} J.,  {Bergin} E.~A.,  {Birnstiel} T.,   {Foreman-Mackey}
  D.,  2018, \mn@doi [\apjl] {10.3847/2041-8213/aac6d7}, \href
  {https://ui.adsabs.harvard.edu/abs/2018ApJ...860L..12T} {860, L12}

\bibitem[\protect\citeauthoryear{{Testi} et~al.,}{{Testi}
  et~al.}{2014}]{testi_2014}
{Testi} L.,  et~al., 2014, in {Beuther} H.,  {Klessen} R.~S.,  {Dullemond}
  C.~P.,   {Henning} T.,  eds, Protostars and Planets VI. p.~339 (\mn@eprint
  {arXiv} {1402.1354}), \mn@doi{10.2458/azu_uapress_9780816531240-ch015}

\bibitem[\protect\citeauthoryear{{Testi} et~al.,}{{Testi}
  et~al.}{2015}]{testi_2015}
{Testi} L.,  et~al., 2015, in Advancing Astrophysics with the Square Kilometre
  Array (AASKA14). p.~117

\bibitem[\protect\citeauthoryear{{Tsukagoshi} et~al.,}{{Tsukagoshi}
  et~al.}{2019}]{tsukagoshi_2019}
{Tsukagoshi} T.,  et~al., 2019, \mn@doi [\apjl] {10.3847/2041-8213/ab224c},
  \href {https://ui.adsabs.harvard.edu/abs/2019ApJ...878L...8T} {878, L8}

\bibitem[\protect\citeauthoryear{{Veronesi}, {Lodato}, {Dipierro}, {Ragusa},
  {Hall}  \& {Price}}{{Veronesi} et~al.}{2019}]{veronesietal2019}
{Veronesi} B.,  {Lodato} G.,  {Dipierro} G.,  {Ragusa} E.,  {Hall} C.,
  {Price} D.~J.,  2019, \mn@doi [\mnras] {10.1093/mnras/stz2384}, \href
  {https://ui.adsabs.harvard.edu/abs/2019MNRAS.489.3758V} {489, 3758}

\bibitem[\protect\citeauthoryear{{Vigan} et~al.,}{{Vigan}
  et~al.}{2017}]{vigan_2017}
{Vigan} A.,  et~al., 2017, \mn@doi [\aap] {10.1051/0004-6361/201630133}, \href
  {https://ui.adsabs.harvard.edu/abs/2017A&A...603A...3V} {603, A3}

\bibitem[\protect\citeauthoryear{{Villenave} et~al.,}{{Villenave}
  et~al.}{2020}]{villenave_2020}
{Villenave} M.,  et~al., 2020, arXiv e-prints, \href
  {https://ui.adsabs.harvard.edu/abs/2020arXiv200806518V} {p. arXiv:2008.06518}

\bibitem[\protect\citeauthoryear{{Walsh}, {Juh{\'a}sz}, {Meeus}, {Dent},
  {Maud}, {Aikawa}, {Millar}  \& {Nomura}}{{Walsh} et~al.}{2016}]{walsh_2016}
{Walsh} C.,  {Juh{\'a}sz} A.,  {Meeus} G.,  {Dent} W. R.~F.,  {Maud} L.~T.,
  {Aikawa} Y.,  {Millar} T.~J.,   {Nomura} H.,  2016, \mn@doi [\apj]
  {10.3847/0004-637X/831/2/200}, \href
  {https://ui.adsabs.harvard.edu/abs/2016ApJ...831..200W} {831, 200}

\bibitem[\protect\citeauthoryear{{Weidenschilling}}{{Weidenschilling}}{1977}]{weidenschilling_1977}
{Weidenschilling} S.~J.,  1977, \mn@doi [\mnras] {10.1093/mnras/180.1.57},
  \href {https://ui.adsabs.harvard.edu/abs/1977MNRAS.180...57W} {180, 57}

\bibitem[\protect\citeauthoryear{{Youdin} \& {Goodman}}{{Youdin} \&
  {Goodman}}{2005}]{youdin_2005}
{Youdin} A.~N.,  {Goodman} J.,  2005, \mn@doi [\apj] {10.1086/426895}, \href
  {https://ui.adsabs.harvard.edu/abs/2005ApJ...620..459Y} {620, 459}

\bibitem[\protect\citeauthoryear{{Zhang}, {Blake}  \& {Bergin}}{{Zhang}
  et~al.}{2015}]{Zhang2015}
{Zhang} K.,  {Blake} G.~A.,   {Bergin} E.~A.,  2015, \mn@doi [\apjl]
  {10.1088/2041-8205/806/1/L7}, \href
  {https://ui.adsabs.harvard.edu/abs/2015ApJ...806L...7Z} {806, L7}

\bibitem[\protect\citeauthoryear{{Zhang}, {Bergin}, {Blake}, {Cleeves},
  {Hogerheijde}, {Salinas}  \& {Schwarz}}{{Zhang} et~al.}{2016}]{zhang_2016}
{Zhang} K.,  {Bergin} E.~A.,  {Blake} G.~A.,  {Cleeves} L.~I.,  {Hogerheijde}
  M.,  {Salinas} V.,   {Schwarz} K.~R.,  2016, \mn@doi [\apjl]
  {10.3847/2041-8205/818/1/L16}, \href
  {https://ui.adsabs.harvard.edu/abs/2016ApJ...818L..16Z} {818, L16}

\bibitem[\protect\citeauthoryear{{Zhang} et~al.,}{{Zhang}
  et~al.}{2018}]{zhang_2018}
{Zhang} S.,  et~al., 2018, \mn@doi [\apjl] {10.3847/2041-8213/aaf744}, \href
  {https://ui.adsabs.harvard.edu/abs/2018ApJ...869L..47Z} {869, L47}

\bibitem[\protect\citeauthoryear{{Zhu} et~al.,}{{Zhu} et~al.}{2019}]{zhu_2019}
{Zhu} Z.,  et~al., 2019, \mn@doi [\apjl] {10.3847/2041-8213/ab1f8c}, \href
  {https://ui.adsabs.harvard.edu/abs/2019ApJ...877L..18Z} {877, L18}

\bibitem[\protect\citeauthoryear{{van der Marel}, {Dong}, {di Francesco},
  {Williams}  \& {Tobin}}{{van der Marel} et~al.}{2019}]{vanderMarel2019}
{van der Marel} N.,  {Dong} R.,  {di Francesco} J.,  {Williams} J.~P.,
  {Tobin} J.,  2019, \mn@doi [\apj] {10.3847/1538-4357/aafd31}, \href
  {https://ui.adsabs.harvard.edu/abs/2019ApJ...872..112V} {872, 112}

\makeatother
\end{thebibliography}

\appendix

\section{Efficiency of radial drift for different sized grains}
\label{sec:drift}

\begin{figure}
    \centering
    \includegraphics[width=\columnwidth]{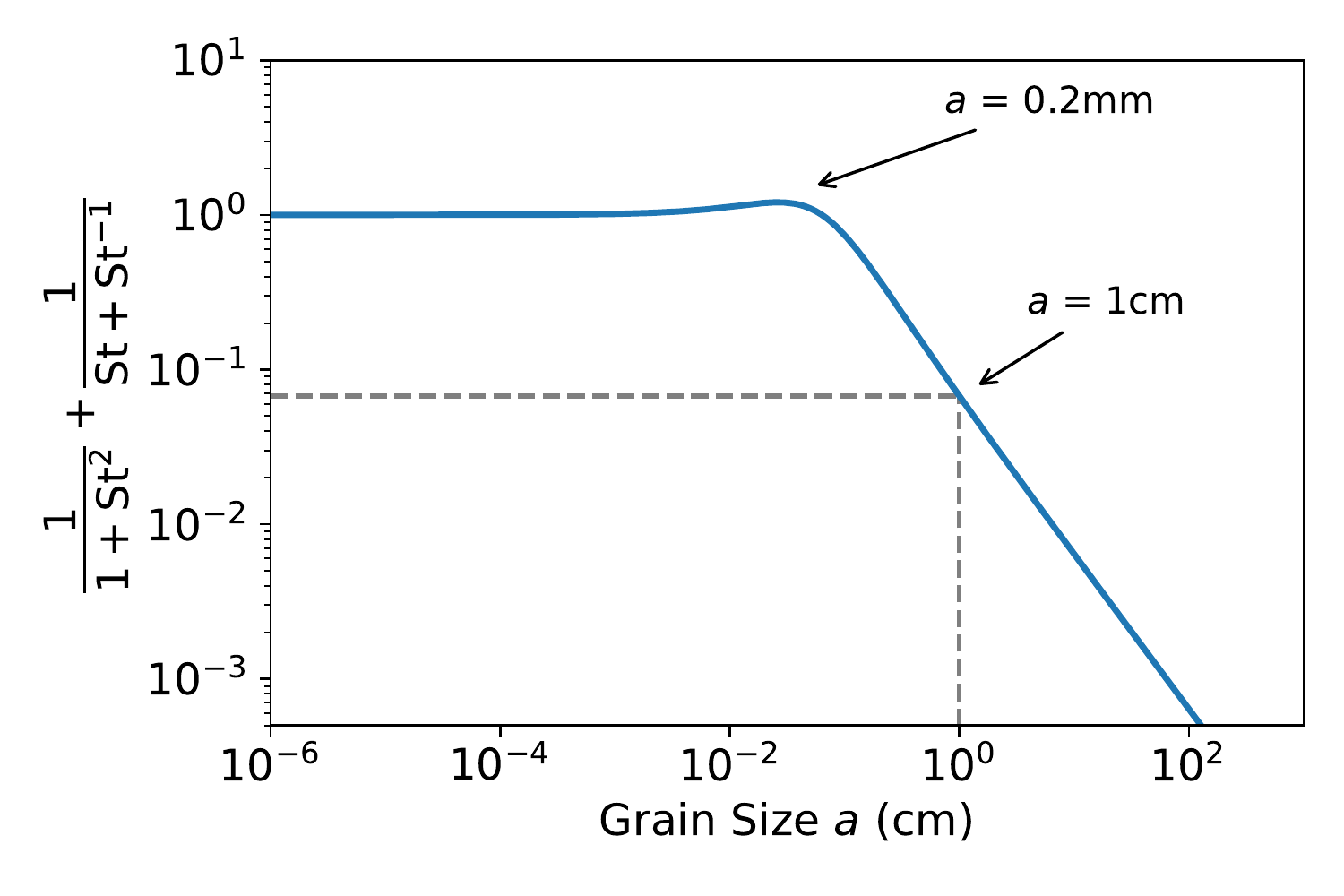}
    \caption{Plot of dimensionless prefactor in Equation \ref{eq:drift} which demonstrates efficency of radial motion of dust, where $St$ was calculated assuming $\rho_s=1$ g cm$^{-3}$ and $\Sigma_\mathrm{g}=0.1$ g cm$^{-2}$.}
    \label{fig:stokes_plot}
\end{figure}

Figure \ref{fig:stokes_plot} shows the efficiency of radial drift of dust in a disc from Equation \ref{eq:drift} for different grain sizes $a$.  We calculated $St$ assuming $\rho_s=1$\,g\,cm$^{-3}$ and $\Sigma_\mathrm{g}=0.1$\,g\,cm$^{-2}$. Dust is perfectly coupled to the gas for $St\ll 1$, optimum trapping occurs for $St=1$ at $a=0.2$\,mm, and as grain sizes become larger, they are increasingly decoupled from the gas resulting in closer to Keplerian orbits.  While the 1\,cm grains studied in this work are not close to the highest efficiency, they still undergo significant amounts of trapping to produce substructure in the disc in the form of rings and gaps (see Figure \ref{fig:hydro}). 

\bsp	
\label{lastpage}
\end{document}